\documentclass[aps, twocolumn, amsmath,amssymb, prx, superscriptaddress]{revtex4-2}
\usepackage{svg,times,upgreek}
\usepackage{color}
\usepackage{graphicx}
\usepackage{dcolumn}
\usepackage{bm}
\usepackage{physics}
\usepackage{hyperref}
\usepackage{url}
\usepackage{amsmath}
\usepackage{tikz-cd}
\usepackage{notes2bib}
\usepackage{ dsfont }
\usepackage{svg}
\usepackage{verbatim}
\usepackage{amssymb}
\usepackage{amsfonts}              
\usepackage{amsthm}                
\usepackage{mathtools}
\usepackage{qcircuit}
\usepackage[normalem]{ulem}
\usepackage{xcolor}

\DeclarePairedDelimiterX{\barpair}[2]{(}{)}{%
  #1\;\delimsize\|\;#2%
}
\theoremstyle{definition}
\newtheorem{theorem}{Theorem}
\newtheorem{prop}[theorem]{Proposition}

\newtheorem{lemma}[theorem]{Lemma}
\newtheorem{definition}[theorem]{Definition}
\newtheorem{corollary}[theorem]{Corollary}

\newcommand{\STr}{\mf{Tr}}
\newcommand{\Ad}{\text{Ad}}
\newcommand{\id}{\text{id}}
\newcommand{\mf}[1]{\mathfrak{#1}}
\newcommand{\cent}[1]{S_\alpha^\diamond\left(#1\right)}
\newcommand{\rent}[1]{S_\alpha\left(#1\right)}
\newcommand{\dent}[1]{S_\alpha^{\diamond\diamond}\left(#1\right)}
\newcommand{\mcal}[1]{\mathcal{#1}}
\newcommand{\renyi}{R\'{e}nyi }
\newcommand{\subc}{\tilde{\mf{C}}}
\newcommand{\jami}{Jamio{\l}kowski }
\newcommand{\mds}[1]{\mathds{#1}}
\newcommand{\supp}[1]{\text{supp}\left(#1\right)}
\newcommand{\maxes}[1]{\mf{c}\left(#1\right)}

\bibnotesetup{note-name    = ,use-sort-key = false} 
\begin{document}

\title{Delocalized and Dynamical Catalytic Randomness and Information Flow}

\author{Seok Hyung Lie}
\author{Hyunseok Jeong}
\affiliation{%
 Department of Physics and Astronomy, Seoul National University, Seoul, 151-742, Korea
}%

\date{\today}

\begin{abstract}

We generalize the theory of catalytic quantum randomness to delocalized and dynamical settings. Our result is twofold. First, we expand the resource theory of randomness (RTR) by calculating the amount of (R\'{e}nyi) entropy catalytically extractable from a correlated or dynamical randomness source. In doing so, we show that no entropy can be catalytically extracted when one cannot implement local projective measurement on randomness source without altering its state.  The RTR, as an archetype of the `concave' resource theory, is complementary to the convex resource theories in which the amount of randomness required to erase the resource is a resource measure. As an application, we prove that quantum operation cannot be hidden in correlation between two parties without using randomness, which is the dynamical generalization of the no-hiding theorem. On the other hand, we study the physical properties of information flow. Popularized quotes like ``information is physical'' by Landauer or ``it from bit'' by Wheeler suggest the matter-like picture of information that can travel from one place to another with the definite direction while leaving detectable traces on its region of departure. To examine the validity of this picture, we focus on that catalysis of randomness models directional flow of information with the distinguished source and recipient. We show that classical information can always spread from its source without altering its source or its surrounding context, like an immaterial entity, while quantum information cannot. Using the framework developed in this work, we suggest an approach to formal definition of semantic quantum information and claim that utilizing semantic information is equivalent to using a partially depleted information source. By doing so, we unify the utilization of semantic and non-semantic quantum information and conclude that one can always extract more information from an incompletely depleted classical randomness source, but it is not possible for quantum randomness sources. 
\end{abstract}

\pacs{Valid PACS appear here}
\maketitle
\section{Introduction} \label{sec:intro}
Flow of information is a key criterion that decides which processes are allowed and which are not in physical theories. For example, there are ostensibly faster-than-light phenomena such as phase velocity (or even group velocity \cite{alexeev2002measurement}) of electromagnetic wave, expansion velocity of far galaxies due to Hubble's law \cite{kaya2011hubble} and collapse of wave function shared between space-like regions, but they are not forbidden by relativity because it is widely considered that those phenomena are not accompanied by faster-than-light propagation of information \cite{diener1996superluminal}. Moreover, oftentimes it is said that nothing can escape black holes, but black holes evaporate by emitting Hawking radiation. A common justification of this is that Hawking radiation does not convey information of objects fallen into the black hole. These examples suggest that information flow is not only as real as flow of any matter as Landauer said ``information is physical," but also has enough independency that warrants focus for its own.

However, what is information, exactly? How is it different from other materialistic entities? Can information propagate from its source to a target without visiting any other regions like a particle, or must it spread to multiple regions like wave? Although we intuitively have vague idea about what information is, answering this question in a universally satisfactory way is highly difficult  considering the sheer vastness of information science. The advent of quantum information theory burdens the already complicated the field of information science with more mystery, and makes us ask the same questions for quantum information.

Quantum information is frequently identified with quantum state and displacement of a quantum state is interpreted as an information flow, but this approach is unsatisfactory since it is not quantum state \textit{per se}, but the variance of quantum state by some information source is what carries information. This observation asks for a dynamical approach to information flow, namely, that identifies information flow with a quantum channel with nonzero capacity, which has been taken in studies on localizable and causal quantum operations \cite{beckman2001causal}.

While largely successful, the picture of information as a varying quantum state and the resultant measurement outcome change treats quantum systems merely as a medium for communication of classical information and overlooks the nature of `quantum information' itself.
Treating pure quantum states informative is contradictory with the perspective of the Shannon information theory \cite{shannon1948mathematical}, where information is identified with randomness. Especially, considering state-dependent restrictions on causality in recent proposals for black hole information paradox such as the Hayden-Preskill protocol \cite{hayden2007black}, the necessity for investigating (semi-)causality in the (partially) static setting is growing lately. Interpreting randomness as information provides a picture that can satisfactorily describe information localized in a region of spacetime and its propagation, as one can assign entropy to each region from their quantum state.

These two perspectives on information are complementary to each other: Randomness of quantum state represents the internal information, or information \textit{inside} a quantum system, and the current state of a quantum system represents the external information, or information \textit{one has} about the system. The latter is often too implicit and heavily depends on the context, hence it is hard to locate and quantify. On the contrary, advantage of internal information is that it is easy to locate and track its presence and propagation. Therefore, to model the directional (quantum) information flow from a source to a unique target, we employ the theory of catalytic quantum randomness and generalize it further to a broader class of randomness sources such as correlated and dynamical sources.

The resource theory, a framework in which a certain physical aspect is abstracted as a resource to analyze the property in question systematically, has been immensely successful in quantum physics and quantum information science. A resource theory identifies resourceful objects (states, operations, etc.) by defining what is considered \textit{free}, meaning that it is easy to perform or prepare, and treating everything that is not free as resourceful. There are many examples of properties for which resource theoretical approach was successful; entanglement \cite{horodecki2009quantum}, coherence \cite{streltsov2017colloquium}, non-Gaussianity \cite{weedbrook2012gaussian}, and many more. These generic resource theories have one thing in common. They are either \textit{convex} or admit convexification. Note that a resource theory is convex when the set of free objects is convex.

The convexity condition is considered natural in many cases; in many recent works \cite{regula2019characterizing, regula2021fundamental, tan2021fisher} on unified approach to resource theory with resource-independent methods, it is assumed that the free set is convex. A common justification is that simply forgetting information, a common method of physically implementing convex sum, cannot generate useful resources. However, this assumption is by no means always justified. Indeed, there are non-convex resource theories such as that of correlation. Statistically mixing two states without correlation can generate correlation, and especially, since the convex hull of the set of all states without correlation is the whole quantum state set, the theory does not allow convexification to form a meaningful resource theory. 

\begin{figure}[t]
    \includegraphics[width=.5\textwidth]{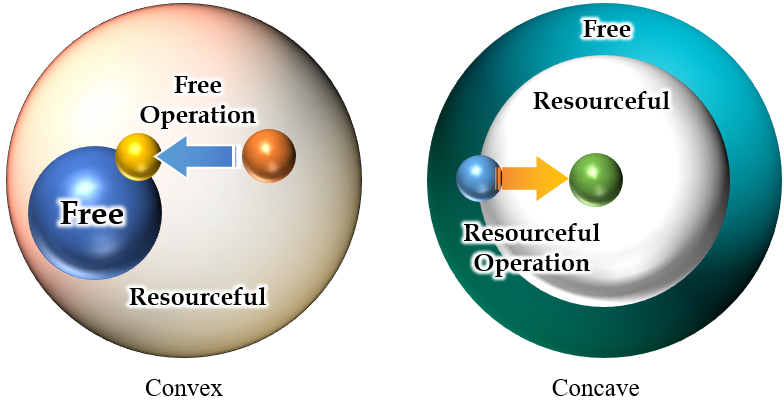}
    \caption{Comparison of convex and concave resource theories. In a convex resource theory, a statistical mixture of two free object is still free, and the action of free operation can only draw a resourceful object closer to the set of free objects. However, in a `concave' resource theory, any statistical mixture of two resourceful object is resourceful, and there is no universal `resource destroying operation'. However, there are resourceful operations that never makes a resourceful object free.}\label{fig:comparison}
\end{figure}

More extremely, there are resource theories that are what we will say to be \textit{concave}. In these resource theories, the set of resourceful objects, not the free objects, is convex (see FIG. \ref{fig:comparison}). In this situation, forgetting information has not only a potential to create resources, but also can never eliminate resources.

The premise that destruction of information is resourceful is natural in both fundamental and practical contexts. Fundamentally, the time evolution of a closed quantum system is given by unitary operations which are invertible, thus it is often said that no quantum information is genuinely destructible (following the usual `state = information' definition). This is the very reason behind the long-lasting controversy on what will happen eventually to quantum information fallen into black holes \cite{polchinski2017black}. Practically, in some cryptographic settings where mutually distrustful participants are interacting, it is impossible for one participant to persuade other participants that some information was deleted from one's data storage without some special assumptions. (It is ridiculous to say ``Hey, I just flipped a coin and I forgot the outcome. Let's bet on which side the coin was." over text message.) This is why one needs a special protocol for coin flipping by telephone \cite{blum1983coin} and more generally cryptographic primitives such as bit-commitment and oblivious transfer.

 Randomness represents both presence and absence of information depending on perspective. The more random an information source is, the less information one already has about the source, equivalently, the more information the soure can yield. Hence, in a sense, forgetting information could create randomness. Thus, an archetype of concave resource theory is the resource theory of randomness (RTR) \cite{boes2018catalytic,lie2019unconditionally,lie2020minent,lie2020randomness,lie2020uniform,lie2021correlational}. In the RTR, pure states are considered free and unitary operations are free operations, but none of them have convex structure. Moreover, there is no universally resource-destroying map \cite{chitambar2019quantum} since every locally randomness-decreasing map should increase randomness globally \cite{lie2021correlational}. On the other hand, the set of mixed states and the set of unital maps, which are considered resourceful in the RTR, are both convex.
 
 Previously, in the RTR, only static and local quantum states with nonzero entropy were considered as randomness sources, but in real life dynamical or global randomness sources are commonplace. Most symbolically, secret key randomly generated and shared by multiple agents is an example of delocalized randomness source, and the simple action of rolling dice itself is a dynamical source of randomness. In this work, we extend the limit of the RTR to encompass utilization of delocalized and dynamical randomness sources by employing the Choi-\jami isomorphism \cite{choi1975completely,jamiolkowski1972linear} and the language of dynamical resource theory \cite{gour2020dynamical}.

In Section \ref{subs:rai}, we argue that the semantics-independent quantitative aspect of information is captured by randomness, and that using no physical properties other than information of an information carrier means not leaking information to the carrier. This motivates the study of catalysis or randomness.

In Section \ref{subs:catran}, we review the resource theory of static and local catalytic randomness , the concept of catalytic decomposition of quantum state, and the catalytic quantum entropies. In Section \ref{subs:delc}, we generalize the theory to the delocalized setting, where multiple parties share a multipartite quantum state as a randomness source and use it to transform another multipartite quantum state. We show that partially-classical structure of multipartite states provides the definition of the delocalized catalytic decomposition and subsequently that of the delocalized catalytic entropies. We show that a Hilbert space can be categorized into two types. A multipartite state is sensitive to the action of unital quantum channels in Type I subspaces, hence no further randomness utilization is possible in there, but any unital channel can be freely applied to type II subspaces without altering the state, so they can yield the quantum advantage of catalytic quantum randomness. In Section \ref{subs:dyca}, we translate the results to the setting of dynamical randomness in which quantum channels are utilized as randomness sources through the Choi-\jami isomorphism. In Section \ref{sec:examples}, we introduce various examples of bipartite states and quantum channels with their catalytic entropies.

In Section \ref{sec:disc}, we discuss the physicality of quantum information and define semantic information in terms of catalytic randomness.

\section{Preliminaries}
\subsection{Notations} \label{subsec:notation}
Without loss of generality, we sometimes identify the Hilbert space $H_X$ corresponding to a quantum system $X$ with the system itself and use the same symbol $X$ to denote both. For any system $X$, $X'$ is a copy of $X$ with the same dimension, i.e., $|X|=|X'|$. When there are many systems other than a system $X$, then all the systems other than $X$ are denoted by $\bar{X}$. However, the trivial Hilbert space will be identified with the field of complex numbers and will be denoted by $\mds{C}.$  We will denote the dimension of $X$ by $|X|$. The identity operator on system $X$ is denoted by $\mds{1}_X$ and the maximally mixed state is denoted by $\pi_X=|X|^{-1}\mds{1}_X$. For any Hermitian matrix $\sigma$, $\lambda_i(\sigma)$ denotes its $i$-th largest eigenvalue including degeneracy, i.e., it is possible that $\lambda_i(\sigma)=\lambda_{i+1}(\sigma)$.  For any Hilbert spaces $X$ and $Y$, $X\leq Y$ denotes that $X$ is a subspace of $Y$. The space of all bounded operators acting on system $X$ is denoted by $\mf{B}(X)$, the real space of all Hermitian matrices on system $X$ by $\mf{H}(X)$. The set of all unitary operators in $\mf{B}(X)$ is denoted by $\mf{U}(X)$. For any matrix $M$, $M^T$ is its transpose with respect to some fixed basis, and for any $M\in\mf{B}(X\otimes Y)$, the partial transpose on system $X$ is denoted by $M^{T_X}$. For any $M\in \mf{B}(X)$, we let $\Ad_M\in\mf{L}(X)$ be $$\Ad_M(K):=MKM^\dag.$$

The space of all linear maps from $\mf{B}(X)$ to $\mf{B}(Y)$ is denoted by $\mf{L}(X, Y)=\mf{B}(\mf{B}(X),\mf{B}(Y))$ and we will used the shorthand notation $\mf{L}(X):=\mf{L}(X, X)$. The set of all quantum states on system $X$ by $\mf{S}(X)$ and the set of all quantum channels (completely positive and trace-preserving linear maps) from system $X$ to $Y$ by $\mf{C}(X , Y)$ with $\mf{C}(X):=\mf{C}(X , X)$. Similarly we denote the set of all quantum subchannels (completely positive trace non-increasing linear maps) by $\tilde{\mf{C}}(X,Y)$ and $\tilde{\mf{C}}(X):=\tilde{{\mf{C}}}(X,X)$. We denote the identity map on system $X$ by $\id_X$. Let $\mcal{T}:M\mapsto M^T$ be the transpose map, and $\dag:M\mapsto M^\dag$ be the adjoint map. For any $\mcal{N}\in\mf{L}(X,Y)$, we define its adjoint $\mcal{N}^\dag(G)$ so that $\langle \mcal{N}^\dag(G),H\rangle=\langle G,\mcal{N}(H)\rangle$ for every $G\in\mf{B}(Y)$  and $H\in\mf{B}(X)$. We define the transpose $\mcal{N}^T(H):=(\mcal{N}^\dag(H^*))^*$, where $G^*$ is the complex conjugation of $G$.

$J_{XX'}^\mcal{N}$ is the Choi matrix of $\mcal{N} \in \mf{L}(X)$ defined as $J_{XX'}^\mcal{N}:=\mcal{N}_X(\phi^+_{XX'})$ where $\phi^+_{XX'}=\dyad
{\phi^+}_{XX'}$ is a maximally entangled state with $\ket{\phi^+}_{XX'}=|X|^{-1/2}\sum_i\ket{ii}_{XX'}$. The mapping $J:\mf{L}(X)\to\mf{B}(X\otimes X')$ defined as $J(\mcal{M}):=J_{XX'}^\mcal{M}$ itself is called the Choi-Jamio{\l}kowski isomorphism \cite{choi1975completely,jamiolkowski1972linear}. We call a linear map from $\mf{L}(X)$ to $\mf{L}(Y)$ a \textit{supermap} from $X$ to $Y$ and denote the space of supermaps from $X$ to $Y$ by $\mf{SL}(X,Y)$ and let $\mf{SL}(X):=\mf{SL}(X,X)$. Supermaps preserving quantum channels even when it only acts on a part of multipartite quantum channels are called \textit{superchannel} \cite{chiribella2008transforming,gour2019comparison,chiribella2009theoretical,burniston2020necessary,bisio2019theoretical,chiribella2013quantum,gour2020dynamical} and the set of all superchannels from $X$ to $Y$ is denoted by $\mf{SC}(X,Y)$ and we let $\mf{SC}(X):=\mf{SC}(X,X)$.   We say a superchannel $\mcal{V}\in\mf{SC}(X)$ is \textit{superunitary} if there are $U_0$ and $U_1$ in $\mf{U}(X)$ such that $\mcal{V}(\mcal{N})=\Ad_{U_1}\circ\mcal{N}\circ\Ad_{U_0}$ for all $\mcal{N}\in \mf{L}(X)$.

 The \textit{supertrace} \cite{lie2021supertrace} is the superchannel counterpart of the trace operation modelling the loss of dynamical quantum information, denoted by $\mf{Tr}$. The supertrace is defined in such a way that the following diagram is commutative: 
\begin{equation}
    \begin{tikzcd}[baseline=\the\dimexpr\fontdimen22\textfont2\relax]
    \mf{L}(X)\ar[d,"J"]\ar[r,"\mf{Tr}"] & \mds{C} \ar[d,"\text{id}_\mds{C}"] \\
    \mf{B}(X\otimes X')\ar[r,"\Tr"]& \mds{C}
  \end{tikzcd}.
\end{equation}
Here, we slightly abused the notations by identifying isomorphic trivial Hilbert spaces $\mds{C}^*\approx\mds{C}\approx\mf{L}(\mds{C})\approx\mf{B}(\mds{C}\otimes \mds{C})$ and letting $J:\mf{L}(\mds{C})\to\mf{B}(\mds{C}\otimes\mds{C})$ be identified with $\text{id}_\mds{C}$. Explicitly,
\begin{equation} \label{eqn:supertrace}
    \mf{Tr}[\mcal{M}]:=\Tr[J_{XX'}^\mcal{M}]=\Tr[\mcal{M}(\pi_X)],
\end{equation}
for all $\mcal{M}\in\mf{L}(X)$. From (\ref{eqn:supertrace}), it is evident why the supertrace corresponds to the loss of information of quantum channels as it is operationally equivalent to the loss of input state (as the input state is assumed to be maximally mixed) and the loss of output state (as the output state is traced out). Similarly to partial trace, $\STr_X$ is a shorthand expression of $\STr_X \otimes \mf{id}_{\bar{X}}$, where $\mf{id}_Y:=\id_{\mf{L}(Y)}$. Note that the supertrace lacks a few tracial properties such as cyclicity, i.e., $\STr[\mcal{A}\circ\mcal{B}]\neq\STr[\mcal{B}\circ\mcal{A}]$ in general, however, it generalizes the operational aspect of trace as the discarding action. For example, for every quantum channel $\mcal{N}$ is normalized in supertrace, i.e, $\STr[\mcal{N}]=1$.

In a similar way, we define the `Choi map' $\mds{J}[\Theta]\in\mf{L}(X\otimes X',Y\otimes Y')$ of supermap $\Theta\in\mf{SL}(X,Y)$ in such a way that the following diagram is commutative:
\begin{equation}
    \begin{tikzcd}[baseline=\the\dimexpr\fontdimen22\textfont2\relax]
    \mf{L}(X)\ar[d,"J"]\ar[r,"\Theta"] & \mf{L}(Y) \ar[d,"J"] \\
    \mf{B}(X\otimes X')\ar[r,"{\mds{J}[\Theta]}"]& \mf{B}(Y\otimes Y')
  \end{tikzcd}.
\end{equation}

Throughout the paper, the direct sum symbol $\oplus$ for operators has two meanings: If $A_i$ are already in the same space and mutually orthogonal, then $\bigoplus_i A_i$ emphasizes such fact and it means simply $\sum_i A_i$. If $B_i$ are not necessarily mutually orthogonal, or even repeated for different $i$, then $\bigoplus_i B_i$ embeds the operators into a larger Hilbert space and make them mutually orthogonal. One possible implementation is $\bigoplus_i B_i := \sum_i \dyad{i}\otimes B_i$. 

\subsection{Superselection rule and $C^*$-algebra}
It is customary to model a quantum state of system $X$ with a density matrix $\rho$ in $\mf{B}(X)$, but it is not necessary to assume that a quantum system has access to all of the full matrix algebra $\mf{B}(X)$. In general, a quantum system can be modelled with a $C^*$-algebra \cite{landsman1998lecture, haagerup2011factorization}, and a finite dimensional $C^*$-algebra is isomorphic to a direct sum of full matrix algebras by the Artin-Wedderburn theorem \cite{artin1927theorie,wedderburn1908hypercomplex}. In other words, for every finite dimensional $C^*$-algebra $\mcal{C}$, there exist finite dimensional Hilbert spaces $X_i$ such that $\mcal{C}\approx \bigoplus_{i=1}^n \mf{B}(X_i)$. 

In fact, it is equivalent to saying that the system $X$ is under \textit{superselection rules} which means that there exists subspaces $\{X_i\}$ of $X$ called the \textit{superselection sectors} such that $\mf{S}(X)\subseteq\bigoplus_i \mf{B}(X_i)$. Therefore, one can interpret that, at least for finite dimensional cases, a $C^*$-algebra $\mcal{C}\approx\bigoplus_{i=1}^n \mf{B}(X_i)$ represents a classical-quantum hybrid system in which a classical information $i$ is not allowed to be in superposition. We call the vector $(|X_1|,|X_2|,\cdots,|X_n|)$ the \textit{dimension vector} of $\mcal{C}$ and $n$ the \textit{dimension rank} of $\mcal{C}$. To make the dimension vector unique, we assume that $|X_1|\geq|X_2|\geq\cdots$ unless there is a pre-defined order of $X_i$ in the given context. When the dimension rank is larger than 1, we say that $\mcal{C}$ is partially classical When $|X_i|=1$ for every $i$, we say that $\mcal{C}$ is (completely) classical. If the dimension rank is 1, we say that $\mcal{C}$ is (totally) quantum.

Remember that $\rho_{AB}$ is called a classical-quantum(C-Q) state when $\rho_{AB}$ can be embedded into the tensor product of $C^*$-algebras $\mcal{C}\otimes\mcal{D}$ where $\mcal{C}$ is classical, i.e., there is a basis $\{\ket{i}_A\}$ of $A$ such that $\rho_{AB}$ has the form
    \begin{equation} \label{eqn:CQ}
        \rho_{AB}=\sum_i p_i \dyad{i}_A\otimes \rho_B^i,
    \end{equation}
for some probability distribution $\{p_i\}$ and quantum states $\rho_B^i\in\mf{S}(B)$. When the roles of $A$ and $B$ are switched, we call it Q-C, and if $\rho_{AB}$ is neither C-Q nor Q-C, then it is called Q-Q. As a generalization, we will call $\rho_{AB}$ partially classical-quantum (PC-Q) if $\rho_{AB}$ can be embedded into the tensor product of $C^*$-algebras $\mcal{C}\otimes\mcal{D}$ where $\mcal{C}$ is partially classical, i.e., there exists a projective measurement $\{\Pi_i\}_{i=1}^n$ with $n>1$ on $A$ ($\Pi_i\Pi_j=\delta_{ij}\Pi_i$ and $\sum_i \Pi_i =\mds{1}_A$) that leaves $\rho_{AB}$ unperturbed. In other words,
    \begin{equation} \label{eqn:PCQ}
        \rho_{AB}=\sum_i(\Pi_i\otimes \mds{1}_B)\rho_{AB}(\Pi_i\otimes \mds{1}_B).
    \end{equation}
 If (\ref{eqn:PCQ}) holds, we also say that $\rho_{AB}$ is generalized block-diagonal with respect to $A=\bigoplus_i A_i$ where $A_i=\supp{\Pi_i}$ \cite{chen2014schmidt}. If the roles of $A$ and $B$ are reversed, we will call it Q-PC. If a bipartite state is both PC-Q and Q-PC, then it is called PC-PC. On the other hand, if a system is not partially classical, we will say that it is totally quantum(TQ), so that a bipartite system that is not PC-Q is now called TQ-Q. One can similarly define Q-TQ, PC-TQ, TQ-TQ states, etc.

\section{Characterizations of catalytic randomness} \label{subs:rai}
In this Section, we give an intuitive motivation for the study of resource theory of catalytic randomness. See Ref. \cite{lie2021correlational} for related discussion.
\subsection{Randomness and Information: Internal and external views}
There is one interpretation of information which considers that information \textit{we have} about systems is the information those systems carry. This kind of interpretation requires or implicitly assumes a user outside of a system, hence we will call it \textit{external information} of the system. From this perspective, randomness of a state is a \textit{noise}. This is why when a pure state becomes mixed, often it is said that information is destroyed \cite{suter2016colloquium}. Similarly, this is why often the no-cloning theorem is interpreted to forbid copying quantum information \cite{dieks1982communication,wootters1982single}, when the exact statement is that it is impossible to copy an arbitrary single pure state. In this sense, a certain aspect of external information of a system can be quantified with nonuniformity \cite{gour2015resource}. However, it actually quantifies the \textit{capability} of carrying information rather than the amount of information \textit{per se}. The external perspective often implicitly assumes implications of a certain piece of information has about other systems, say, $n$-photon state $\ket{n}$ carries more energy than vacuum state $\ket{0}$. However, this meaning heavily depends on its user and hence is highly subjective.

In this framework, a state only represents the current status of a system, and its \textit{change} is considered to carry information in this interpretation. It requires sender's coding and receiver's decoding, thus external information tends to be more dynamical. In other words, one says that (external) information at region $A$ \textit{does not} flow to region $B$ via a map $\Lambda_{A\to B}$ when $\Lambda_{A\to B}$ is constant, i.e.,
\begin{equation}
    \Lambda_{A\to B}(\rho)=\Lambda_{A\to B}(\sigma),
\end{equation}
for every state $\rho$ and $\sigma$. If it is not the case, one says that information flows from $A$ to $B$. Information source that provides information to be encoded is often treated implicitly and assumed to be outside of information transmission processes.

However, there is another line of thought on information that focuses on information contained \textit{inside} a system, or the \textit{internal information}. For example, a cylinder filled with gas can be said to contain a lot of information as one can learn a lot of data by inspecting the configuration of its constituting gas molecules. Simply put, internal information is information of a system when treated as a black box. The Shannon information theory is built on the observation that acquisition of the state of a system is considered more informative when the state appears more random before the acquisition. Hence, classically, the \textit{information content} or \textit{surprisal} $I(x)$ of an event $x\in \mcal{X}$ is defined as \cite{shannon1948mathematical,mcmahon2007quantum}
\begin{equation} \label{eqn:surprisal}
    I(x)=-\log_2\Pr[X=x],
\end{equation}
so that the \textit{average} information content, or the \textit{Shannon entropy} of probability distribution $P$ is
\begin{equation}
    H(P)=-\sum_{x\in\mcal{X}} P(x) \log_2 P(x).
\end{equation}
The von Neumann entropy of a quantum state $\rho$ defined as
\begin{equation} \label{eqn:von}
    S(\rho)=-\Tr[\rho\log_2\rho],
\end{equation}
can be interpreted in the same fashion, so that $S(\rho)$ represents the amount of classical internal information of $\rho$. (We will elaborate on the meaning of `classical' afterwards.) Internal information perspective treats information explicitly, for example, since one can calculate the entropy of each local system, it is easy to locate and quantify information. From this perspective, randomness and information are identified, and maximally mixed states are maximally informative states. Since the state completely decides internal information of a system, the role of observer or context is minimal in this interpretation. 

From this perspective, correlation is formed when information propagates from its source to other systems. Hence, when system $A$ and $B$ initially prepared in an uncorrelated state $\sigma_A\otimes \rho_B$ interact, one says that (internal) information does not flow from $A$ to $B$ if, for any extension $\sigma_{AR}$ with some reference system $R$, systems $RB$ are still uncorrelated after the interaction. If not, information propagates from $A$ to $B$ through the interaction.

These two interpretations look completely contradictory to each other, however, they are actually two complementary views on information. For example, classically, one way to measure external information is the \textit{relative entropy} $D(P\|U)$ from the maximally uniform distribution $U(x)=|\mcal{X}|^{-1}$, where $D(P\|Q)$ is the relative entropy that measures the statistical separation between two distributions and is given as $D(P\|Q):=\sum_{x\in\mcal{X}} P(x)\log_2(P(x)/Q(x))$. They are in the following clear-cut trade-off relation, 
\begin{equation}
    H(P) + D(P\|U) = \log_2 |\mcal{X}|,
\end{equation}
for the case of the von Neumann entropy the same thing holds \textit{mutatis mutandis}, hence discussion about information inside or about a system are essentially the same except for their opposite signs up to additive constant.

Moreover, two notions of information flows introduced above are actually equivalent to each other \cite{lie2021correlational}; if internal information does not flow, then neither does external information. Therefore, to treat information flow on the same footing with any other flow of physical entities, we will first characterize flow of internal information and try to explain all the other informational phenomena in terms of internal information. This is in line with relational approach to quantum mechanics by Rovelli \cite{rovelli1996relational} and Everett \cite{everett2015relative}. To treat the noisy aspect and the informational aspect of randomness neutrally, we will use `randomness' and `information' interchangeably so that all the results can be used regardless of one's interpretation of randomness.

In this context, an information source stripped of its semantic meaning is nothing but a  randomness source. Hence, we can say that randomness captures the universal quantitative aspect of information independent of their meaning, and Shannon information theory successfully quantifies this non-semantic information with entropic quantities. Thus, in this work, we will use the term `randomness' to emphasize this semantics-independent quantitative aspect of internal information. This is what referred to as `Information-B' among three types of information in the \textit{Handbook of Philosophy of Information} \cite{adriaans2008philosophy}. (By Ref. \cite{adriaans2008philosophy}, `Information-A' focuses on semantics, and `Information-C' focuses on algorithmic complexity.) We will focus on the analysis of non-semantic information first, but we will tackle the problem of analyzing semantic information in Section \ref{subsec:sem}.

\subsection{Catalytic randomness and information flow}

In Introduction, we observed that information can be localized and displaced, and takes an important role in physical theory, sometimes even more important than ostensible material entities. Hence, it is natural to treat information as a physical entity that a system can possess and to identify its properties.

How is information different from other physical entities? First of all, for information to be physically relevant, it should leave detectable effects on its receiver, however, not every detectable change is made by information. If someone breaks your window by throwing a rock to notify you, is it information in the rock that broke the window? It is natural to conclude that information exchange merely accompanied the event and it is the kinetic energy of the rock that broke the window. Like this example, in general, exchange of information is mixed up with other physical effects.

\begin{figure}[t]
    \includegraphics[width=.4\textwidth]{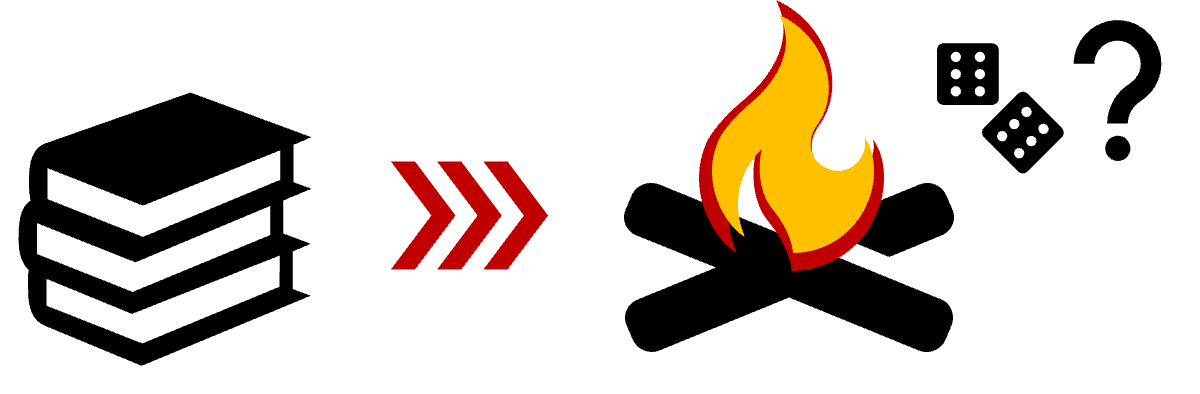}
    \caption{A book is a randomness (or information) source, but not every usage of it is pure randomness utilization. For example, it is hard to say that burning a book utilizes only the randomness of the book, as it leaves evidently detectable physical traces on it. Intuitively it is clear that any usage of a book that necessitates non-negligible physical alternation of the book is not a pure information utilization. Therefore we claim that (pure) randomness utilization must not leave any locally detectable statistical change on the randomness source.}\label{fig:fire}
\end{figure}

What would a `pure' information source that does not yield any physical resources other than information look like? For this to be possible, no detectable change of physical resource in the source is allowed, therefore its state should stay unchanged. It means that no detectable change can be caused by the other system it is interacting with, equivalently, there is no information flow from it into the source. We could say that this kind of interactions have \textit{directional information flow} in which information only flows from a distinguished information source to its user and not the other way around. This is the process we may call a purely information utilizing process and we claim that it must satisfy the following mutually related criteria (See FIG \ref{fig:fire}).

\begin{enumerate}
    \item \textit{Random} : The state of an information source must be random to be informative.
    
    \item \textit{Correlating} : After a use of an information source, it forms correlation with its user, altering their global state.
    
    \item \textit{Directional} : Information flows from an information source to its user exclusively, not the other way around.
\end{enumerate}

We already discussed why randomness is crucial for an information source. Information usage is entropy extraction process, hence correlation between a source and it user is naturally built in the process and the amount of correlation formed can be interpreted as the amount of randomness extracted from the source \cite{lie2021correlational}.

Directionality criterion can be applied both on fundamental and various practical levels. A person may not be able to read a book leaving absolutely no traces (e.g. not perturbing molecular arrays of the book at all), but if the trace is `practically' (whatever that means in a given context) undetectable so that its statistical state is left unchanged, then we consider that the person only used the information content of the book on that practicality level. This fact allows us to circumvent the question of fundamental nature of randomness in light of deterministic time evolution of classical/quantum mechanics in closed systems, as there are events appear random on practical level regardless of the underlying law of nature.

For example, even when one interacts with a cylinder filled with gas without altering any thermodynamic parameters such as temperature and volume, another person who memorized all the configurations of molecules of the gas is able to detect the change. However, to that person, the gas was not random from the beginning. For a person to whom only the macroscopic quantities of the gas were known, the gas can still appear intact. If a randomness source behaves the same way in every statistical aspect after an interaction, we consider it unaffected.

Hence, in a purely information (or randomness) utilizing process, the information carrier simply enters the interaction and leaves it while staying in the same quantum state. Nevertheless, the information carrier could cause changes of other systems. This fits the definition of catalysis and the carrier can be considered a catalyst. This is one of the main reasons why the study on catalysis of randomness is motivated. Nonetheless, we intuitively know that information itself can be `depleted' for individual users \cite{lie2021correlational}. For example, a novel is no longer interesting once a reader finishes reading it and remembers all the plot despite the fact that the book is physically unchanged. This can be explained by the correlation built between the carrier and the user, which is a purely informational quantity. On the other hand, the memory of the reader initially prepared in a pure state becomes random after forming correlation with other systems. Hence correlation-forming can be interpreted as randomness extraction. These two observations motivate the study of a theory that sounds contradictory on the surface level, the resource theory of catalytic randomness.

In this work, we will investigate the properties of quantum information flow by studying catalytic quantum randomness. One may claim that this type of `noninvasiveness' is a characteristic of classical randomness and should not be required from quantum randomness, because of the inherent perturbing nature of quantum measurement. However, such a claim comes from confusing quantum information with quantum state. The latter contains every physical description of a quantum system, be it informational or not, and we are trying to characterize the former in this work. Indeed, one cannot interact nontrivially with a quantum system in a pure state without perturbing it, but a system with zero entropy has no information to provide in the first place. Therefore, a quantum information source must be in a mixed state, and we know that we can extract information, measured by entropy, without perturbing the mixed state \cite{boes2018catalytic, lie2019unconditionally, lie2020randomness, lie2020uniform, lie2021correlational}.

Note that we do not concern ourselves with the mechanism of \textit{randomness generation}. Just as resource theory of entanglement cares more about manipulation of already existing entanglement rather than studying the protocol of entanglement establishment (which is different from entanglement distillation), resource theory of randomness is more about utilization of pre-existing randomness sources regardless of their generation mechanism. Hence, `quantum randomness (source)' in this work is not related to what conventionally referred to as quantum randomness, which usually means a classical random variable generated by measuring a quantum system, stored in classical memory. Quantum randomness in this work means the randomness of quantum systems enjoying its quantum coherence, represented by mixed quantum states. This is the reason why one need not answer the question of `what is the true origin of randomness?' before using the resource theory of randomness, as users with different criteria for randomness can still use the same theory.

\section{Resource Theory of Randomness}

\subsection{Catalytic randomness} \label{subs:catran}
In this Section, we summarize and review the results of the correlational resource theory of catalytic randomness \cite{lie2021correlational}. Suppose that $A$ is allowed to borrow a system $B$ called \textit{catalyst} in the quantum state $\sigma_B$ to implement a quantum channel $\mcal{N}$. $A$ is allowed to interact with $B$ but should return the system $B$ in its original state $\sigma_B$ after every interaction. This can be summarized as the following two conditions. When a bipartite unitary $U$ on systems $A$ and $B$ is used to implement a quantum channel $\rho \mapsto \mcal{N}(\rho)$ with a catalyst $\sigma$ for arbitrary possible input state $\rho$, i.e.

\begin{equation} \label{eqn:catal1}
    \Tr_B \Ad_U(\rho_{A} \otimes \sigma_{B}) = \mcal{N}(\rho), \quad\forall \rho\in\mf{S}(A).
\end{equation}
The catalyst $\sigma$ should retain its original randomness, i.e. spectrum, after the interaction regardless of the input state $\rho$, i.e.
\begin{equation} \label{eqn:catal2}
    \Tr_A \Ad_U(\rho_{A} \otimes \sigma_{B}) = \sigma_{B} \quad\forall \rho\in\mf{S}(A).
\end{equation}
 The conditions above require the catalyst to be insensitive to dynamically changing state of the target system. This dynamical definition can be re-expressed in the Heisenberg picture and in the static setting; we can require the catalyst to be insensitive to the change of action on the target system.
\begin{theorem} \label{thm:picinv}
    Condition (\ref{eqn:catal2}) is equivalent to any of the following.
    
    $(i)$ For some state $\rho_A \in \mf{S}(A)$ and for every superchannel $\Theta\in\mf{SC}(A)$, the transformed bipartite quantum channel $(\Theta_{A}\otimes \mf{id}_B)(\Ad_U)$ fixes the marginal state $\sigma_B$, i.e.
\begin{equation} \label{eqn:cacon}
    \Tr_A[(\Theta_{A}\otimes \mf{id}_B)(\Ad_U)(\rho_A\otimes \sigma_B)]=\sigma_B.
\end{equation}

    $(ii)$ When $\rho_A\in\mf{S}(A)$ is given, for any ancillary system $R$, a unitary operator $U\in\mf{V}(RA)$ and the state given as $\tau_{RA}=\Ad_V(\dyad{0}_R\otimes\rho_{A})$, the following holds.
\begin{equation} \label{eqn:cacon2}
    \Tr_A[\id_R\otimes\Ad_U(\tau_{RA}\otimes\sigma_B)]=\tau_R\otimes\sigma_B^{(V)}.
\end{equation}
    Here, the marginal state $\sigma_B^{(V)}$ may depend on $V$.
\end{theorem}
A more detailed discussion on the condition given in terms of superchannels can be found in Section \ref{subsec:sem}.

We can see that one-way constraint on information flow is picture-invariant, i.e., independent of the interpretation of randomness; Condition $(i)$ requires that system $B$ is indifferent to the change of dynamical process on $A$. Condition $(ii)$ requires that no internal information of $A$, held by $R$, is leaked to $B$. Therefore, we can use whichever picture that suits the given situation to simplify expressions and unless specified otherwise, we will consider catalysis of randomness in the form of (\ref{eqn:catal1}) and (\ref{eqn:catal2}). 

 The possible dependence of $\sigma_B^{(V)}$ on the process $V$ hints that Condition $(ii)$ only prohibits leakage of internal information. However, there is actually no external information leakage, because if there are two unitary operators $V_1$ and $V_2$ that leads to different $\sigma_B^{(V)}$, then by preparing an additional ancillary qubit prepared in $\ket{+}$ state making it control which operator among $V_i$ is applied on $RA$, one can contradict Condition $(ii)$. Moreover, by Stinespring dilation, one can easily see that unitary operation $\Ad_V$ in Condition $(ii)$ can be replaced by any quantum channel. These observations combined yield Condition $(iii)$ in the next Proposition, and also a completely static characterization, Condition $(iv)$. Considering the Choi-\jami isomorphism, Condition $(iv)$ being equivalent to $(i)$ is evident.

\begin{prop} \label{prop:morequ}
    Conditions in Theorem \ref{thm:picinv} are equivalent to the following conditions.
    
    $(iii)$ When $\rho_A\in\mf{S}(A)$ is given, for any quantum channel $\mcal{N}\in\mf{C}(A,RA)$ with $\tau_{RA}:=\mcal{N}(\sigma_A)$, we have
    \begin{equation}
        \Tr_A[\id_R\otimes\Ad_U(\tau_{RA}\otimes\sigma_B)]=\tau_R\otimes\sigma_B.
    \end{equation}
    
    $(iv)$ For any quantum $\rho_{RA}$ state whose marginal state $\rho_A$ is full-rank, we have
    \begin{equation}
        \Tr_A[\id_R\otimes \Ad_U(\rho_{RA}\otimes\sigma_B)]=\rho_R\otimes\sigma_B.
    \end{equation}
\end{prop}
The approach of Condition $(iii)$ that treats the initial setup, the subsequent interaction and the partial trace out as a superchannel that maps interjected quantum channel into an outcome state is akin to the approach of Modi \cite{modi2012operational} for dynamics of non-Markovian open quantum systems. The requirement of full-rankedness of $\rho_A$ in Condition $(iv)$ is rather technical than physical, as the set of full-rank states is dense in the set of all states. However precisely one prepares a quantum state, there could be an infinitesimal noise in the process that renders the prepared state full-rank.

Although the catalyst changes by some unitary operator $V$, any unitary operator can be reverted by a deterministic agent and it is intuitive that randomness of quantum state only depends on its spectrum, so we accept this definition. We will call the bipartite interaction described in (\ref{eqn:catal1}) and (\ref{eqn:catal2})  a \textit{catalysis} or a \textit{catalysis process} and a quantum channel that can be implemented by catalysis a \textit{catalytic} quantum map or channel. For example, the quantum channel $\mcal{N}$ in (\ref{eqn:catal1}) is catalytic. We will call the bipartite unitary operator used for catalysis a \textit{catalysis unitary} operator.

We will say that $U$ is compatible with $\sigma$ (and vice versa) if (\ref{eqn:catal2}) holds. If (\ref{eqn:catal2}) holds with the right hand side replaced with $V\sigma_B V^\dag$ with some unitary operator $V$ on $B$, then they are said to be compatible up to local unitary. Using an incompatible catalyst for a given catalysis unitary operator will lead to change of the catalyst after the interaction. For the sake of convenience, we will often use the definition of the compatibility for the cases where $\sigma_B$ is an unnormalized Hermitian operator, too. Similar randomness-utilizing processes were considered in previous works, under the name noisy operations \cite{horodecki2003reversible, scharlau2018quantum, gour2015resource} or thermal operations. However, most studies were focused on the implementation of the transition between two fixed quantum states and the existence of a feasible catalyst for that task. Here, we are more interested in the implementation of quantum channel, independently of potential input state, with a given catalyst. However, later we will see that this characterization is also relevant to state transitions, too. In the following Theorem, we review the characterization of catalytic unitary operators and compatibility.

\begin{theorem}[\cite{lie2021correlational}] \label{thm:catuni}
    A bipartite unitary operator $U$ acting on system $AB$ is catalytic if and only if $U^{T_B}$ is also unitary. Also, a catalytic unitary operator $U$ is compatible with $\sigma_B$ if and only $[U,\mds{1}_A\otimes
    \sigma_B]=0$.
\end{theorem}

\begin{figure}[t]
    \includegraphics[width=.5\textwidth]{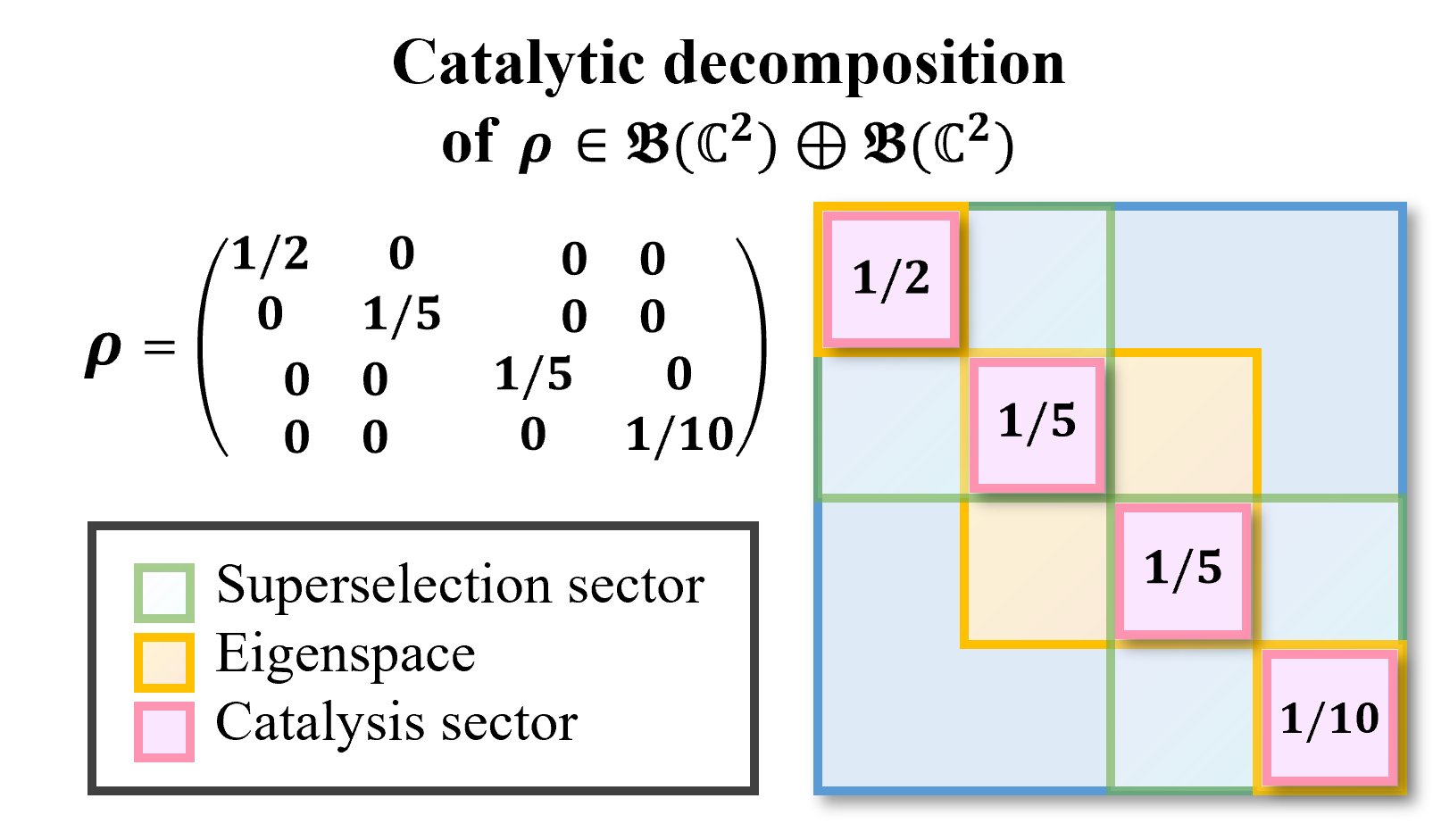}
    \caption{Catalytic decomposition of a density matrix. A superselection rule forbids between subspaces called superselection sectors, and each density matrix has an eigenspace for each distinct eigenvalue. The intersection of a superselection sector and an eigenspace is called a catalysis sector and it plays an important role in calculating the catalytic entropies.}\label{fig:decomp}
\end{figure}

Unlike in resource theories with resource-destroying maps, in the RTR, convertibility between randomness sources is not a very interesting problem since they are either too trivial or too restrictive. Any two quantum states are freely interconvertible if and only if they share the spectrum. If we expand to conversions under catalytic maps, then the problem becomes trivial again since between any two quantum states $\rho \succ \sigma$, there exists a random unitary operation $\mcal{F}$ , which is also catalytic, such that $\mcal{F}(\rho)=\sigma$ \cite{watrous2018theory}. Therefore, focusing on how much and what kind of randomness is required to implement certain tasks is much more important than merely asking if the conversion exists.

Now we turn to the problem of quantifying the amount of resource one can extract from a source. The amount of information extracted can be quantified with the mutual information
$$I(A:B)=S(A)+S(B)-S(AB),$$
between $A$ and $B$. However, under the catalysis constraints, the local state of $B$ is invariant and the entropy of global state is invariant, i.e., $S(AB)=S(\rho_A)+S(\sigma_B)$, hence the mutual information after catalysis is equal to the entropy change of system $A$, i.e., $\Delta I(A:B)=S(\mcal{N}(\rho_A))-S(\rho_A)$. Therefore, we will count the entropy increase as the amount of extracted resource during catalysis of quantum randomness. This interpretation is consistent with the view that treats randomness as noise. Generalizing this, we interpret that randomness gained through catalytic maps is from the influx of information. Thus, although there is no simple generalization of mutual information for R\'{e}nyi entropies, we will also use the \renyi entropies to measure the extracted information from a randomness source.

It was shown in Ref. \cite{lie2020uniform,lie2021correlational} that non-degeneracy of eigenvalues of a mixed state restricts catalysis of quantum randomness. Accordingly, \textit{the catalytic \renyi entropy} $\cent{\sigma}$ of order $\alpha\geq0$ of an arbitrary quantum state $\sigma\in\mf{S}(X)$ can be calculated from its spectral decomposition. By spectral decomposition, we mean $\sigma=\sum_i\lambda_i\Pi_i$ with eigenvalues $\lambda_i$ of $\sigma$. Here, we require $\Pi_i\Pi_j=\delta_{ij}\Pi_i$, $\sum_i\lambda_i r_i=1$ and the injective mapping $i\mapsto \lambda_i\geq0$. If there are superselection rules imposed on $X$, i.e. $\mf{S}(X)\subseteq\bigoplus_i \mf{B}(X_i)$ for some mutually orthogonal subspaces $X_i$ of $X$, then we require instead that $\supp{\Pi_i}\leq X_{f(i)}$ for some unique subspace of $B$, $X_{f(i)}$ and that $i\mapsto (\lambda_i,X_{f(i)})$ is injective.  We denote the rank of each block by $r_i:=\Tr[\Pi_i]$. Let the spectral decomposition satisfying these requirements be called the \textit{catalytic decomposition} of a quantum state and we call each $\supp{\Pi_i}$ a catalysis sector of $\sigma$ (see FIG.\ref{fig:decomp}).

In this sense, a catalyst compatible with a catalytic unitary operator could be considered a partially classical quantum system only whose classical information (the weight of each catalysis sector) is known.

For any $\sigma$ with the catalytic decomposition $\sigma=\sum_i \lambda_i \Pi_i$, define a density matrix $\maxes{\sigma}$ given as
\begin{equation}
    \maxes{\sigma}=\bigoplus_i \frac{\lambda_i}{r_i}\mds{1}_{r_i^2},
\end{equation}
where $\mds{1}_{r_i^2}=\text{diag}(1,\cdots,1)$ is the identity matrix of size $r_i^2$. It was shown in Ref.\cite{lie2021correlational} that any mixed state catalytically transformed from a pure state by using randomness source $\sigma$ majorizes $\maxes{\sigma}$ and catalytic transformation into $\maxes{\sigma}$ from a pure state is also achievable. In other words, $\maxes{\sigma}$ is the most random state that can be catalytically created with $\sigma$ from a pure state. Let us call $\maxes{\sigma}$ the randomness-exhausting output (REO) of $\sigma$. Since every \renyi entropy is Schur-concave, and the maximum (global) entropy production of a quantum channel is achieved with a pure state input \cite{lie2021correlational}, $\rent{\maxes{\sigma}}$ is the the maximum \renyi entropy catalytically extractable from randomness source $\sigma$, and we call it the catalytic \renyi entropy $\cent{\sigma}$ of $\sigma$. $\cent{\sigma}$ has the following explicit expression in terms of the catalytic decomposition of $\sigma$.

\begin{equation} \label{eqn:catent}
    \cent{\sigma}:=\frac{1}{1-\alpha}\log_2\left[\sum_i \lambda_i^\alpha r_i^{2-\alpha}\right].
\end{equation}
The important extreme cases are the catalytic von Neumann entropy $\lim_{\alpha \to 1}\cent{\sigma}=S^\diamond(\sigma):=-\sum_i \lambda_ir_i \log_2 (\lambda_i/r_i)$, the min-catalytic entropy $\lim_{\alpha\to\infty}\cent{\sigma}=S_{\min{}}^\diamond(\sigma):=-\log_2\left[\max_i \lambda_i/r_i\right]$, and the max-catalytic entropy $\lim_{\alpha\to 0+}\cent{\sigma}=S_{\max{}}^\diamond(\sigma):=\log_2\left[\sum_ir_i^2\right]$. The catalytic entropies are important because of the following operational meaning.
\begin{theorem}[\cite{lie2021correlational}] \label{thm:catent}
    The maximum amount of catalytically extractable \renyi entropy of order $\alpha\geq0$ from a randomness source $\sigma$ is its catalytic \renyi entropy defined as $\cent{\sigma}$.
\end{theorem}

Although it is known that, for a given quantum channel, more entropy is produced on a purification than on a mixed state, it could be still cumbersome to find an input state that yields the maximum entropy production for a given channel. However, if our intention is to check if the channel produces entropy at all, then the following Proposition says that inputting a maximally entangled state is enough. See Appendix for proof.

\begin{prop}\label{prop:nogen}
    A catalytic map cannot generate randomness with any input state if and only if it cannot produce randomness by acting on a part of a maximally entangled state.
\end{prop}

\subsection{Delocalized catalytic randomness} \label{subs:delc}

In the last Section, we only considered randomness sources that are in isolation from other systems. In this Section, we generalize catalysis of randomness to correlated randomness sources. The necessity of such a generalization naturally arises when multiple parties share correlated data to implement some delocalized information processing task. There are abundant examples of correlated randomness source. Multiple copies of the same book are all correlated and altering one copy can be physically detected when the copies are compared. People also share secret keys to encrypt another shared data by using it. Oftentimes, one does not only use the information of the system they are directly in contact with, but also utilize its relation with the outer world. One may also only have access to small part of large system but still want to restrict the information flow into the whole system.

Correlated information sources are also generic in the quantum setting, too. Treating systems correlated with a given information source not explicitly could cause huge confusion, as it was exemplified in the controversy around M{\o}lmer's conjecture \cite{molmer1997optical}. A way to resolve the confusion is explicitly take account of the correlation, especially the entanglement, between laser light and the laser device. A detailed discussion can be found in Appendix.

The detailed setting of delocalized catalysis of randomness is as follows. Instead of one party, let there be two parties, Alice $(A_0)$ and Alex $(A_1)$, separated in different laboratories. They start with an initial bipartite state $\rho_{A_0A_1}$, and they are provided with a bipartite state $\sigma_{B_0B_1}$ as a randomness source that they should return unchanged. Alice can only control $A_0B_0$ and Alex can only control $A_1B_1$. They try to transform their initial state into some other state $\mcal{N}(\rho_{A_0A_1})$ without altering the randomness source. We allow no communication between them in this process because communication establishes new shared randomness sources between them.

In the quantum setting, Alice will apply unitary operator $U_0$ to $A_0B_0$, and Alex will apply $U_1$ to $A_1B_1$.  Just like the original catalysis scenario, they are required to preserve $\sigma_{B_0B_1}$ after the interaction, regardless of their initial state $\rho_{A_0A_1}$. This requirement can be summarized as
\begin{equation} \label{eqn:dlc1}
    \Tr_{B_0B_1}[\Ad_{U_0\otimes U_1}(\rho_{A_0A_1}\otimes\sigma_{B_0B_1})]=\mcal{N}(\rho_{A_0A_1}),
\end{equation}
with some quantum channel $\mcal{N}\in\mf{C}(A_0A_1)$ and
\begin{equation} \label{eqn:dlc2}
    \Tr_{A_0A_1}[\Ad_{U_0\otimes U_1}(\rho_{A_0A_1}\otimes\sigma_{B_0B_1})]=\sigma_{B_0B_1},
\end{equation}
for all $\rho_{A_0A_1}\in \mf{S}(A_0\otimes A_1)$. We will call this type of catalysis a \textit{delocalized catalysis} of randomness and when it is needed to emphasize it, we call $\sigma_{B_0B_1}$ in this situation the \textit{delocalized randomness source}. We say that the catalysis unitary operator pair $(U_0,U_1)$ is compatible with $\sigma_{B_0B_1}$ if (\ref{eqn:dlc2}) holds, and vice versa, and we say that they are compatible up to local unitary when there exists some $V_i\in\mf{U}(B_i)$ for $i=0,1$ such that (\ref{eqn:dlc2}) holds with the right hand side substituted with $\Ad_{V_0\otimes V_1}(\sigma_{B_0B_1})$. If we need to emphasize, we will call the special case $V_i=\mds{1}_{A_i}$ for $i=0,1$ the \textit{canonical} case. When we focus on the action of each local party, we say that $U\in\mf{U}(A_0B_0)$ is compatible with $\sigma_{B_0B_1}$ on $B_0$ when $(U,\mds{1}_{A_1B_1})$ is compatible with $\sigma_{B_0B_1}$.

We can observe that delocalized catalysis can be considered a special case of catalysis of randomness. Thus, Theorem \ref{thm:catuni} applies here too, hence $U_0\otimes U_1$ must be catalytic, implying that $U_0$ and $U_1$ must be catalytic unitary operators themselves. Also, for $\sigma_{B_0B_1}$ to be compatible with $U_0\otimes U_1$, it must be that $[U_0\otimes U_1,\sigma_{B_0B_1}]=0$. In local catalysis of randomness, a randomness source cannot yield randomness if and only if it is a pure state. Does the same result hold in delocalized catalysis too?

Now, we observe that, in delocalized catalysis, each party can only interact locally with their shared randomness source without altering the global state of it. Considering that no communication between them is allowed, we could guess that each of them must leave the correlated source intact, independent of each other's action. What is the condition for this to be possible? It was recently proved that if a subsystem is not even partially classical, meaning that no nontrivial projective measurement can be implemented on its local system, then the quantum state shared with it is sensitive to changes caused by unital quantum channels \cite{lie2021faithful}.
\begin{lemma}[\cite{lie2021faithful}.] \label{lem:sensitive}
    For any quantum state $\rho_{AB}$, $(\mcal{N}_A\otimes \id_B) (\rho_{AB})\neq \rho_{AB}$ for any unital channel $\mcal{N}_A\neq\id_A$ if and only if $\rho_{AB}$ is a TQ-Q state.
\end{lemma}
It is because quantum correlation can detect local randomizing disturbance and it hinders the catalytic utilization of the randomness source. From these observations, we can identify the bipartite states that cannot yield randomness and show that there are quantum states that are not pure but unable to provide any randomness catalytically.

\begin{theorem} \label{thm:delodep}
    No randomness can be catalytically extracted from a bipartite quantum state if and only if it is TQ-TQ.
\end{theorem}

 The reason why catalysis sectors were identified in local catalysis of randomness was that they are the maximum subspace within which nontrivial unital channels can be applied in an unconstrained fashion without affecting the state of randomness source. (See FIG.\ref{fig:decomp}.) The same idea can be applied in delocalized catalysis of randomness, and we should identify the maximum subspaces within which local parties can apply unital channels without any constraint and the danger of altering the state of the given randomness source.

At this point, we introduce the concept of essential decomposition, which provides the canonical decomposition of a partially classical system into classically distinguishable sectors (subspaces of the Hilbert space of each local system) for a PC-Q state. In other words, when we say a PC-Q state is `partially classical', we mean that there is a local projective measurement that does not perturb the state, and the essential decomposition identifies what is the maximally informative measurement of such kind.

\begin{figure}[t]
    \includegraphics[width=.4\textwidth]{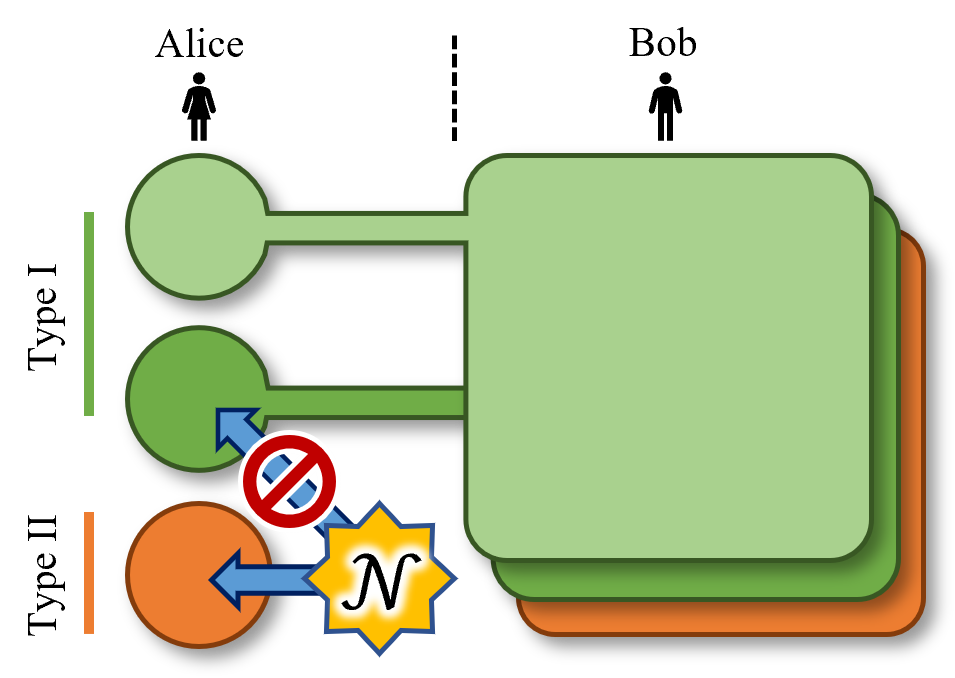}
    \caption{Type I and type II subspaces of the essential decomposition of $A$ (Alice) for $\rho_{AB}$. Alice is quantumly correlated with $B$ (Bob) on Type I subspaces, hence $\rho_{AB}$ resists the action of unital maps without leaving detectable effects on it. But, Alice is uncorrelated with Bob on type II subspaces, so an arbitrary unital channel can be applied to type II subspaces without changing $\rho_{AB}$.}\label{fig:types}
\end{figure}

\begin{definition} \label{def:CPQ}
     Let $\rho_{AB}\in\mf{S}(AB)$ be a bipartite quantum state. $A=\bigoplus_i A_i$ is the \textit{essential decomposition} of $A$ for $\rho_{AB}$, ($\Pi_i:=\mds{1}_{A_i})$ if
     
     $(i)$ For every $i$,
     \begin{equation}
         [\Pi_i\otimes\mds{1}_B,\rho_{AB}]=0.
     \end{equation}
     
     $(ii)$ Each $(\Pi_i\otimes \mds{1}_B)\rho_{AB}(\Pi_i\otimes \mds{1}_B)$ is either a TQ-Q state ($i\in\mcal{I}_A$, ``type I'') or a product state of the form $\pi_{A_i}\otimes\sigma_B$ for some $\sigma\in\mf{S}(B)$ ($i\in\mcal{II}_A$, ``type II'') after normalization.
     
     $(iii)$ Whenever any projector $P$ does not commute with some of $\Pi_i$, we have $[P\otimes\mds{1}_B,\rho_{AB}]\neq0$.

    If none of $\Pi_i$ is the identity operator on $A$, we say $\rho_{AB}$ is a PC-Q state with respect to the essential decomposition $A=\bigoplus_i A_i$.
\end{definition}
We will use the term ``type I (or II)'' for the indices $i$, the corresponding components $(\Ad_{\Pi_i}\otimes\id_B)(\rho_{AB})$ and the subspaces $A_i$ accordingly. The essential decomposition is unique: See Appendix \ref{app:unique} for the discussion on the uniqueness of essential decomposition. We will call the corresponding decomposition of $\rho_{AB}=\sum_i (\Ad_{\Pi_i}\otimes \id_B)(\rho_{AB})$ the essential decomposition of $\rho_{AB}$ on $A$.

Why are type I and type II separated? TQ-Q state are known to be sensitive to the perturbations of unital maps \cite{lie2021faithful}, thus it is impossible to interact through a catalytic unitary operator without leaving detectable effects. Hence, TQ-Q components are separated as type I. Any PC-Q state can be further decomposed into TQ-Q state, but if it is in a product state, then they can yield quantum advantage as we will see soon, thus they are separated as type II. On the other hand, the essential decomposition is related with the structure of entropy non-increasing state under unital channels \cite{zhang2011neumann,hiai2011quantum}, in which there are only two types of components, one which only permits unitary operations (corresponding to type I), and the other which permits any unital subchannel but should be the maximally mixed state (corresponding to type II).

The essential decomposition captures the intuitive idea of `classical sectors' of PC-Q states as the following Theorem shows. It says that any `randomizing transformation' acting on the partially classical part of a PC-Q state, represented by unital maps, that preserves the whole state must respect the classical structure of the partially classical system. Additionally, it says that the unital map can act nontrivially only when there is no correlation in each classical sector.

\begin{theorem} \label{thm:pcqunt}
     A unital channel $\mcal{N}\in\mf{UC}(A)$ fixes a quantum state $\rho_{AB}$ that is PC-Q with respect to the essential decomposition $A=\bigoplus_i A_i$ (let $\Pi_i:=\mds{1}_{A_i}$) with corresponding type index sets $\mcal{I}_A$ and $\mcal{II}_A$ if and only if $\mcal{N}$ preserves every subspace $A_i$ and acts trivially on $A_i$ when $i\in\mcal{I}_A$.
\end{theorem}

See Appendix \ref{app:ess} for a deeper analysis of essential decomposition. Now we introduce a bipartite generalization of catalytic decomposition that we will call the \textit{delocalized catalytic decomposition} through the essential decomposition.
\begin{definition} \label{def:essdec}
    Let $\rho_{AB}$ be a bipartite quantum state with the essential decompositions of $A=\bigoplus_i A_i$ and $B=\bigoplus_i B_i$, with $\Pi_i^A:=\mds{1}_{A_i}$ and $\Pi_j^B:=\mds{1}_{B_j}$. The type index sets for each decomposition are given as $\mcal{I}_A$, $\mcal{II}_A$, $\mcal{I}_B$ and $\mcal{II}_B$, respectively. The delocalized catalytic decomposition (DCD) of a bipartite quantum state $\rho_{AB}\in\mf{S}(AB)$ is the spectral decomposition of the following form,
    \begin{equation} \label{eqn:DCD}
        \rho_{AB}=\bigoplus_{i,j} (\Pi_i^A\otimes\Pi_j^B)\rho_{AB}(\Pi_i^A\otimes\Pi_j^B).
    \end{equation}
\end{definition}
 Since the essential decompositions are unique for $A$ and $B$ respectively, the DCD is also unique for $\rho_{AB}$. This definition is slightly more complicated than the definition of the catalytic decomposition for single-partite systems, but it is required to identify the basic building blocks of a delocalized randomness source. Most notably, each component in the DCD is still compatible with any catalysis unitary operators of the original catalysts, just as every component in the catalytic decomposition of single-partite catalysts is compatible with any catalysis unitary operator compatible with the catalyst before the decomposition. (See Appendix for more information.) This observation leads us to the following definition of the \textit{delocalized catalytic \renyi entropy}.

\begin{definition} \label{def:DCD}
     For the DCD of $\rho_{AB}$ given in (\ref{eqn:DCD}), we let $\tau_i:=\dyad{0}$ if $i\in\mcal{I}_A$ and  let $\tau_i:=\pi_{T_i}$, where $T_i=\mds{C}^{|A_i|^2}$ if $i\in\mcal{II}_A$. Similarly, we let $\kappa_j:=\dyad{0}$ if $j\in\mcal{I}_B$ and  let $\kappa_j:=\pi_{K_j}$, where $K_j=\mds{C}^{|B_j|^2}$ if $j\in\mcal{II}_B$. Also, let $p_{ij}:=\Tr[(\Pi_i^A\otimes\Pi_j^B)\rho_{AB}]$. Then, the delocalized catalytic  \renyi entropy $\dent{\rho_{AB}}$ of $\rho_{AB}$ is defined as the following way.
     \begin{equation}\label{eqn:DCE}
         \dent{\rho_{AB}}:=\rent{\bigoplus_{i,j} p_{ij} \tau_i\otimes\kappa_j}.
     \end{equation}
     Here, we call the state $\mf{d}(\rho_{AB}):=\bigoplus_{i,j} p_{ij} \tau_i\otimes\kappa_j$ the delocalized randomness-exhausting output (DREO) of $\rho_{AB}$.
\end{definition}

Just like the catalytic entropies, the delocalized catalytic entropies also have the same kind of operational meaning.

\begin{theorem} \label{thm:delext}
    The maximum \renyi entropy that can be catalytically extracted from a delocalized randomness source $\sigma_{B_0B_1}$ is its delocalized catalytic \renyi entropy.
\end{theorem}

Hence, we successfully quantified the amount of catalytically extractable randomness in the delocalized setting. This analysis of static but delocalized randomness sources can be directly applied to dynamical randomness sources through the Choi-\jami isomorphism in the next Section.

Note that if there is no correlation in the delocalized randomness source, i.e., $\sigma_{B_0B_1}=\sigma_{B_0}\otimes\sigma_{B_1}$, then there are no type I subspaces in the essential decompositions, so delocalized catalysis simply reduces to two independent local catalyses with $\dent{\sigma_{B_0}\otimes\sigma_{B_1}}=\cent{\sigma_{B_0}}+\cent{\sigma_{B_1}}$.

We remark that multipartite generalization of delocalized catalysis or randomness is straightforward. Each party in delocalized catalysis behave locally and there are no collective maneuvers needed. Hence, the delocalized catalytic decomposition is simply the collection of the essential decomposition of each party, so for an $N$-partite quantum state $\rho_{12\cdots N}$, with each party $X=1,2,\cdots,N$, one can partition the $N$ parties into $X:\bar{X}$ and find the essential decomposition. The rest of procedures, e.g. calculating the catalytic entropies and implementing the catalysis, are immediate once the delocalized catalytic decomposition is found.

\subsection{Dynamical catalytic randomness}\label{subs:dyca}

\begin{figure*}[t]
    \includegraphics[width=.9\textwidth]{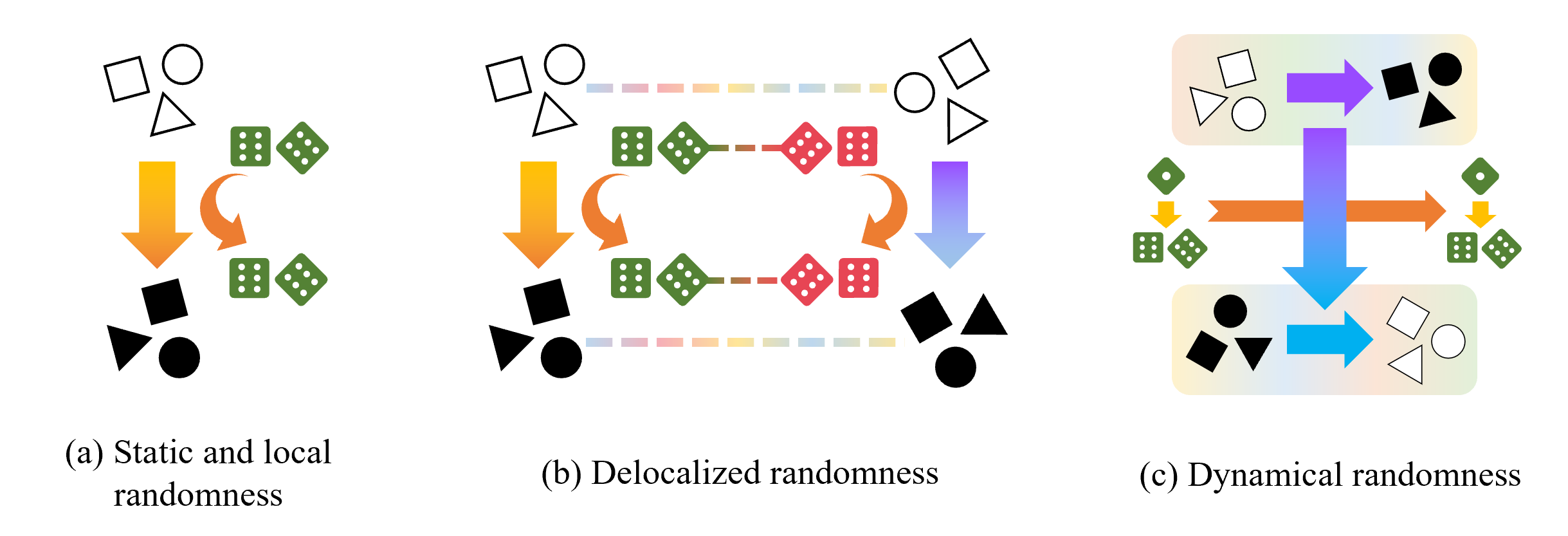}
    \caption{Comparison of three types of catalysis of quantum randomness. Randomness represented by dices enters the interaction and leaves it locally unchanged but correlated with the system. As it can be seen from diagrams (b) and (c), delocalized catalysis and dynamical catalysis of randomness are intimated related; rotating one diagram by 90 degrees makes it very similar to the other one.}\label{fig:three}
\end{figure*}

So far, we have only considered static randomness sources, whose classical examples include random number tables and secret keys. In a more realistic situation, however, \textit{dynamical} sources of randomness are common. For example, when a group of people are playing a tabletop board game, they do not usually play the game with a random number table prepared in advance; they roll a dice to generate randomness on the spot. For example, a record of the result of a previously ($|X|$-faced) dice roll can be modelled by a static state, i.e., the maximally mixed state $\pi_X$, but the action of rolling a dice can be modelled by the depolarizing map $\mcal{R}\in\mf{C}(X)$,
\begin{equation}
    \mcal{R}(\rho)=\pi_X\Tr[\rho],
\end{equation}
for any initial state $\rho$ of the dice with classical system $X$. Even in this case, we claim that catalysis of randomness utilization is still required. In other words, if you have no idea for which game it is used and only observe the dice rolling, then the channel you used as a randomness source must retain its original form. This `information non-leaking' property is very important for characterizing pure randomness utilization \cite{lie2021correlational}, and we require that a randomness source must not remember for which operation it was used and must retain its probabilistic properties regardless of the result of the implemented operation. See Section \ref{subs:rai} for more discussion. This requirement can be formulated as follows.

When one tries to catalytically transform a quantum channel $\mcal{N}$ into $\Theta(\mcal{N})$ by using a quantum channel $\mcal{R}$ as a randomness source, we assume that only applying bipartite unitary operators to input and output systems of $\mcal{N}$ and $\mcal{R}$ is allowed as no randomness producing operation is allowed other than $\mcal{R}$. (See Section \ref{subsec:concave}.)  We will model the complete loss of information about a dynamical quantum process with the \textit{supertrace}, denoted by $\mf{Tr}$, which represents completely losing information on input and output system of a given process, i.e. $\mf{Tr}(\mcal{N}):=\Tr[\mcal{N}(\pi_X)]$. (See Section \ref{subsec:notation}.)

In this work, we will mainly focus on the case where the target channel $\mcal{N}$ and the randomness source channel $\mcal{R}$ act at the same time. In other words, they act on their respective systems \textit{in parallel}. Formally, we say a superchannel $\Theta \in \mf{SL}(A)$ is \textit{catalytic} when there is a bipartite superunitary operation $\Omega \in \mf{L}(AB)$ and a channel $\mcal{R} \in \mf{C}(B)$ such that

\begin{equation} \label{eqn:dycat1}
    \mf{Tr}_B\Omega(\mcal{N} \otimes \mcal{R})= \Theta(\mcal{N}),
\end{equation}
and
\begin{equation} \label{eqn:dycat2}
    \mf{Tr}_A\Omega(\mcal{N} \otimes \mcal{R})= \STr[\mcal{N}]\mcal{R}, 
\end{equation}
for all $\mcal{N}\in\subc(A)$. (See Section \ref{subs:CPmap} for a discussion on the set of $\mcal{N}$.) We will call the whole process a (dynamical) \textit{catalysis} and say that $\mcal{R}$ is used as a \textit{randomness source} (channel) or a \textit{catalyst}. If a superunitary operation $\Omega$ can be used to implement a catalytic superchannel, then it is called a \textit{catalysis superunitary operation}, or it is said to be \textit{catalytic}. A randomness source channel $\mcal{R}$ and a catalysis superunitary operation $\Omega$ is said to be compatible with each other when (\ref{eqn:dycat1}) and (\ref{eqn:dycat2}) hold for some superchannel $\Theta$ and every $\mcal{N}\in\mf{C}(A)$.

Since a superunitary $\Omega$ can be decomposed into the actions of preunitary $U_0$ and postunitary $U_1$ \cite{gour2020dynamical}, i.e., $\Omega(\mcal{N})=\Ad_{U_1}\circ\mcal{N}\circ\Ad_{U_0}$, therefore (\ref{eqn:dycat1}) and (\ref{eqn:dycat2}) can be expressed as $\STr_B[ \Ad_{U_1}\circ\mcal{N}\otimes\mcal{R}\circ\Ad_{U_0}]=\Theta(\mcal{N})$ and $\STr_A[ \Ad_{U_1}\circ\mcal{N}\otimes\mcal{R}\circ\Ad_{U_0}]=\mcal{R}$. By considering the Choi matrices, we get the following expressions
\begin{equation} \label{eqn:equco1}
    \Tr_{BB'}[\Ad_{U_1\otimes U_0^T}(J_{AA'}^\mcal{N}\otimes J_{BB'}^\mcal{R})]=J_{AA'}^{\Theta(\mcal{N})},
\end{equation}
and
\begin{equation} \label{eqn:equco2}
    \Tr_{AA'}[\Ad_{U_1\otimes U_0^T}(J_{AA'}^\mcal{N}\otimes J_{BB'}^\mcal{R})]=J_{BB'}^{\mcal{R}},
\end{equation}
for all $\mcal{N}\in\subc(A)$. Note that every $\rho_{XX'}\in \mf{S}(X\otimes X')$, there exists a $\mcal{M}\in\subc(X)$ such that $J_{XX'}^{\mcal{M}}\propto\rho_{XX'}$, and vice versa. It follows that (\ref{eqn:equco1}) and (\ref{eqn:equco2}) are equivalent to the following requirements, in turn:
\begin{equation} \label{eqn:isoco1}
    \Tr_{BB'}[\Ad_{U_1\otimes U_0^T}(\rho_{AA'}\otimes J_{BB'}^\mcal{R})]=\mds{J}[\Theta](\rho_{AA'}),
\end{equation}
and
\begin{equation} \label{eqn:isoco2}
    \Tr_{AA'}[\Ad_{U_1\otimes U_0^T}(\rho_{AA'}\otimes J_{BB'}^\mcal{R})]=J_{BB'}^{\mcal{R}},
\end{equation}
for every $\rho_{AA'}\in\mf{S}(A\otimes A')$. Here, $U_1$ acts on $AB$ and $U_0^T$ acts on $A'B'$. Now, we can observe that (\ref{eqn:dycat1}) and (\ref{eqn:dycat2}) are only a special case of (\ref{eqn:dlc1}) and (\ref{eqn:dlc2}) after some change of notations, thus we can conclude that $\Omega$ is catalytic if and only if $U_0^{T_A}\otimes U_1^{T_{B'}}$ is unitary. It is equivalent to saying both $U_0$ and $U_1$ are catalytic themselves.

\begin{theorem} \label{thm:supuni}
    A superunitary operation $\Omega:\mcal{N} \mapsto \Ad_{U_1} \circ \mcal{N} \circ \Ad_{U_0}$ is catalytic if and only if both $U_0$ and $U_1$ are catalytic. Also, $\Omega$ is compatible with $\mcal{R}$ if and only if $U_0\otimes U_1^T$ is compatible with $J_{BB'}^\mcal{R}$, i.e., 
    \begin{equation} \label{eqn:vancom2}
        [U_1\otimes U_0^T,\mds{1}_{AA'}\otimes J_{BB'}^\mcal{R}]=0.    
    \end{equation}
\end{theorem}
The vanishing commutator condition (\ref{eqn:vancom2}) follows from Theorem \ref{thm:catuni}. When $\mcal{E}(\rho)=\pi_B\Tr[\rho]$ is the depolarizing map on $B$, its Choi matrix is $J_{BB'}^\mcal{E}=\pi_B\otimes \pi_{B'}$, therefore $[U_1\otimes U_0^T, \mds{1}_{AA'}\otimes J_{BB'}^\mcal{E}]=0$ for any $U_0$ and $U_1$. It implies that, similarly to that every catalysis unitary operator is compatible with the maximally mixed state, every catalysis superunitary operations is compatible with the depolarizing map. In other words, a fair (quantum) dice roll can always provide randomness without leaking information.

There could be many possible measures of randomness extracted from randomness source, but from the formal similarity of static and dynamical catalysis, we will use $\rent{J_{AA'}^{\Theta(\mcal{N})}}-\rent{J_{AA'}^\mathcal{N}}$, for every $\alpha\geq0$, as a measure of extracted randomness. When $\alpha=1$, $\rent{J_{AA'}^\mcal{N}}$ is called the map entropy $S^\text{map}(\mcal{N})$ of channel $\mcal{N}$ \cite{zhang2011neumann,roga2008composition}. Theorem \ref{thm:supuni} immediately yields an upper bound to the amount of randomness catalytically extractable from a randomness source channel $\mcal{R}\in\mf{C}(B)$, namely, $\rent{J_{AA'}^{\Theta(\mcal{N})}}-\rent{J_{AA'}^\mathcal{N}}\leq \cent{J_{BB'}^\mcal{R}}$, where $J_{BB'}^\mcal{R}$ is interpreted to be an element of $\mf{B}(B\otimes B')$ without any superselection rule. However, unitary operators of the form $U_1\otimes U_0^T$ are not of the most general form of 4-partite unitary operator that can act on $AA'BB'$, it is not evident if $\cent{J_{BB'}^\mcal{R}}$ is the maximally extractable \renyi entropy extractable from $\mcal{R}$, counted with the increase of the \renyi entropy of the Choi matrix.

However, from its equivalence with delocalized catalysis of randomness, we can simply use the delocalized catalytic entropies to measure the maximally extractable randomness of arbitrary channel.

\begin{definition}
     The catalytic \renyi entropy $\cent{\mcal{R}}$ of a quantum channel $\mcal{R}\in\mf{C}(B)$ is
     \begin{equation}
         \cent{\mcal{R}}=\dent{J_{BB'}^\mcal{R}}.
     \end{equation}
\end{definition}

The framework of dynamical quantum randomness encompasses the static quantum randomness too. Any static randomness source modelled as a quantum stat $\sigma_B$ can be described as preparation channel $\mcal{P}(\alpha)=\alpha \sigma$ in $\mf{C}(\mds{C},B)$, whose Choi matrix is simply $\mcal{J}_{\mds{C}B}^\mcal{P}=\sigma$, hence $\cent{\mcal{P}}=\dent{\mcal{J}_{\mds{C}B}^\mcal{P}}=\cent{\sigma}$.

We, now, leave a remark on a more general case of catalysis of dynamical quantum randomness. In general, a target channel and a randomness source channel need not be applied simultaneously, and one preceding another is obviously possible. For example, if we assume that the randomness source is applied after the target channel, then we should modify the catalysis conditions as follows. For all $\mcal{N}\in\subc(A)$,

\begin{equation} \label{eqn:gycat1}
    \mf{Tr}_B\,\mcal{U}_3\circ\mcal{R}_B \circ\mcal{U}_2\circ\mcal{N}_A\circ \mcal{U}_1= \Theta(\mcal{N}),
\end{equation}
and
\begin{equation} \label{eqn:gycat2}
     \mf{Tr}_A\,\mcal{U}_3\circ\mcal{R}_B \circ\mcal{U}_2\circ\mcal{N}_A\circ \mcal{U}_1= \mf{Tr}[\mcal{N}]\mcal{R},
\end{equation}
with some superchannel $\Theta \in \mf{SC}(A)$ and some unitary operations $\mcal{U}_i\in\mf{U}(AB)$ for $i=1,2,3$. One can see that the unitary operation $\mcal{U}_2$ in the middle hinders the transforming this process into a delocalized catalysis process. Although we can show that $\mcal{U}_1$ must be a catalytic unitary operation by tracing out both sides of (\ref{eqn:gycat2}), still many other parts of this process is left for further inquiry. Hence, we leave the complete characterization of dynamical catalysis of this type as an open question for the moment. Nonetheless, when there is no randomness in the randomness source $\mcal{R}$, i.e., if $\mcal{R}$ is a unitary process, then one can rump $\mcal{U}_2\circ\mcal{R}_B\circ\mcal{U}_1$ into a single unitary operation, hence it reduces to the dynamical catalysis discussed before, with trivial randomness source, $\id_B$. This fact will be used when we prove the no-stealth theorem in a later section.

\subsection{Partially depleted catalyst and semantic information} \label{subsec:sem}
In previous Sections, we have observed that randomness captures the probabilistic aspect of information that is independent of its semantics. However, the everyday notion of information heavily depends on the semantic properties of information, hence one might find that the discussion of previous Sections misses a large portion of discussion on information. Indeed, the semantic side and the quantitative side of information are notorious for being hard to unify. Nevertheless, in this Section, we venture into the realm of semantic information and attempt to spell out the formalism of semantic information in our framework of catalytic randomness.

Floridi \cite{floridi2013philosophy} defines semantic information as well-formed, meaningful and truthful data. As Shannon's approach to information, which we take in the quantum setting, is probabilistic rather than propositional, we will focus on the `meaningful' part. This definition immediately assumes the existence of reference systems that are related with the carrier of semantic information, as data cannot be meaningful when it is isolated from the outer world. For example, we consider a recipe for some dish meaningful because the recipe is correlated with the properties of the ingredients, which appear random in the Bayesian sense to those who are a novice at cooking. Another example is maps; a map is meaningful compared to any other picture because it corresponds to the geography of the real world.

Therefore, we will try to be value-neutral when it comes to deciding what counts as meaningful and claim that the existence of correlation between information carrier and the object you are going to interact with, the target system, is the key characteristic of semantic information in the context of our formalism. The situation is similar with delocalized catalysis of randomness, but there is an important difference that interaction between information source and target system is allowed and the correlation between the two systems need not be preserved because the target system is now allowed to be altered. Recall that only the state of information source is required to be preserved in our definition of (pure) information utilization.

One of the most typical example is Szilard's engine. Suppose that a gas molecule $G$ in a piston can be either of two states of being in the left half of the piston $\ket{l}_G$ or being in the right half $\ket{r}_G$. Let the molecule be in the maximally mixed state, 
\begin{equation} \label{eqn:szc}
    \rho_G=\frac{1}{2}\dyad{l}_G+\frac{1}{2}\dyad{r}_G.
\end{equation}
A common precondition of Szilard's engine is the acquisition of information about the position of the molecule. Acquisition of information requires the existence of a information carrier that gets correlated with its reference, hence we spell it out as $C$, i.e.,
\begin{equation} \label{eqn:szb}
    \frac{1}{2}\dyad{``l"}_C\otimes\dyad{l}_G+\frac{1}{2}\dyad{``r"}_C\otimes\dyad{r}_G.
\end{equation}
The states $\ket{``l"}_C$ and $\ket{``r"}_C$ are orthogonal to each other and contain the classical information about the state of $G$. By conditioning on the state of $C$, we can initialize the molecule $G$ by applying a reversible process, so that the final state of $CG$ is
\begin{equation} \label{eqn:sza}
    \left(\frac{1}{2}\dyad{``l"}_C+\frac{1}{2}\dyad{``r"}_C\right)\otimes\dyad{r}_G.
\end{equation}
As one can see, we only used the system $C$ as an information source so the state of $C$ is left unaltered but that of $G$ is changed. Observe that the end result is the mere transfer of entropy from $G$ to $C$, which is the key observation needed to solve Maxwell's demon problem.

Our way of modelling semantic information requires two systems, the information source that only provides information and the target system that can be physically affected. If we admit this asymmetry between them, then we need a mathematical characterization of their difference. This distinction is important as Korzybski said ``A map is not the territory''  \cite{korzybski1958science}.

As we have seen in Theorem \ref{thm:picinv}, we could expect that there exist different characterizations of semantic information in each pictures, dynamical (Heisenberg) and static (Schr\"{o}dinger). To construct the dynamical characterization, let us go back to the example of Szilard's engine. When we used the information source, our initial intention was initializing the position of the gas molecule. However, we could always change our mind and do whatever we want with the information we acquired from the source other than initializing the gas molecule into the right half of the cylinder. We claim that this alternation of plan, strictly happening to the action on the target system, must not affect the information source. This requirement, which is a generalization of Condition $(i)$ of Theorem \ref{thm:picinv}, can be expressed concretely as follows.

\begin{definition}[S:A] \label{def:seminf}
We say that a bipartite unitary operation $\mcal{U}=\Ad_U$ with $U\in\mf{U}(AB)$ utilizes \textit{(semantic) information} of $B$ in a bipartite state $\sigma_{AB}$ when for any superchannel $\Theta\in\mf{SC}(A)$, $\mcal{U}_\Theta:=(\Theta_{A}\otimes \mf{id}_B)(\mcal{U})$ does not affect $B$, i.e., there exists $\eta_B\in\mf{S}(B)$ such that for all $\Theta\in\mf{SC}(A)$,
\begin{equation} \label{eqn:semcat}
    \Tr_A[\mcal{U}_\Theta(\sigma_{AB})]=\eta_B.
\end{equation}
\end{definition}
We remark that such $\eta_B$ in (\ref{eqn:semcat}) must be unitarily similar to $\sigma_B$. (See Appendix.) For the static characterization, imagine that we redistribute the information of system $A$ to a larger joint system $RA$ by applying some channel $\mcal{N}_{A\to RA}$. Because of the correlation formed between $R$ and $A$, when static information of $A$ is leaked to $B$ by the interaction between $A$ and $B$, there will be a change in the correlation between $R$ and $B$. Based on this speculation, we can formulate the following Definition in the same spirit with Condition $(iii)$ of Proposition \ref{prop:morequ}.

\begin{definition}[S:B] \label{def:seminf3}
We say that a bipartite unitary operation $\mcal{U}=\Ad_U$ with $U\in\mf{U}(AB)$ utilizes \textit{(semantic) information} of $B$ in a bipartite state $\sigma_{AB}$ when for any state $\tau_{RAB}=(\mcal{N}_{A\to RA}\otimes \id_B)(\sigma_{AB})$ with a quantum channel $\mcal{N}_{A\to RA}$, we have
\begin{equation} \label{eqn:semcat3}
    \Tr_A[(\id_R\otimes\mcal{U})(\tau_{RAB})] = 
    (\id_R\otimes\Ad_{V}) (\tau_{RB}),
\end{equation}
with some $V\in\mf{U}(B)$.
\end{definition}

Alternatively, since we have already developed the definition of using only information of a local system in a multipartite quantum state, one may rather import the definition of delocalized catalysis of randomness and claim the following.

\begin{definition}[S:C] \label{def:seminf4}
We say that a bipartite unitary operation $\mcal{U}=\Ad_U$ with $U\in\mf{U}(AB)$ utilizes \textit{(semantic) information} of $B$ in a bipartite state $\sigma_{AB}$ when $U$ is compatible with $\sigma_{AB}$ on $B$ up to local unitary as a delocalized catalyst.
\end{definition}

The main result of this Section is that these seemingly different definitions of semantic information are equivalent. In other words, utilization of semantic information is fundamentally not different from delocalized catalysis of randomness. Hence, `using only information of system $B$ in correlated systems $ABC\cdots$' can be universally discussed without paying attention to which is allowed to be altered and which system is used as an information source other than $B$. This can be concretely expressed as follows.

\begin{theorem} \label{thm:semthm}
    Definitions (S:A), (S:B) and (S:C) are equivalent.
\end{theorem}

Proof is in Appendix. This result unifies many notions of information usage introduced so far as it will be demonstrated afterwards. So, we will simply drop `semantic' when we refer to this type of information usage. First of all, we can observe that non-semantic (quantum) information is a special case of semantic information by considering uncorrelated $\sigma_{AB}=\sigma_A\otimes\sigma_B$.

Without loss of generality, unless we explicitly state `up to local unitary', we will only consider the `canonical' cases; we assume that no nontrivial unitary operation is applied on $B$ after the interaction for the sake of simplicity.

One can observe that this characterization of semantic information utilization is actually equivalent to catalysis of \textit{partially depleted} randomness source, the characterization of which was an open problem raised in Ref. \cite{lie2021correlational}. It is because now we consider randomness sources that are initially correlated with the target system, and we concluded that randomness sources are consumed by forming correlation with its user. It is in contrast with the previous Sections where randomness sources were assumed to be initially in a product state with the target system. Therefore, we can consider utilization of semantic information is also in the formalism of catalytic quantum randomness.

We already know that a bipartite state $\sigma_{AB}$ that is Q-TQ cannot yield catalytic randomness on $B$. Hence, we get the following Corollary which shows that utilization of semantic quantum information is impossible when you cannot use non-semantic quantum information when you are required not to disturb the information source, just as it is in the classical setting.

\begin{corollary}
     If $\sigma_{AB}$ is Q-TQ, then no non-product bipartite unitary operation can utilize only semantic information of $B$ in $\sigma_{AB}$.
\end{corollary}

An important example of quantum state that is Q-TQ is pure states with full Schmidt rank. Hence, as pure states were not useful for delocalized catalysis of randomness, they also do not allow utilization of pure semantic information. Note that the requirement of full Schmidt rank can be circumvented by limiting the local Hilbert spaces to the support of each marginal state, as they are the only physically relevant Hilbert spaces. 

One may wonder, since utilization of information of $B$ in $\sigma_{AB}$ allows information flow from $A$ to $AB$ and from $AB$ to $B$, if it is possible to circumvent the restriction of one-way information flow by breaking the process in two steps so that one has net flow of information from $A$ to $B$. Indeed, even if $M$ and $N$ are catalytic unitary operators compatible with $\sigma_B$, the same need not hold for their composition $NM$.

However, such circumvention is impossible after all; one lesson we learned from the observations of previous Sections is that one should be explicit about reference systems when one treats information from the internal information perspective. First of all, if system $A$ starts from the maximally mixed state uncorrelated with any other systems, then the action of arbitrary catalytic unitary compatible with the state of $B$ does not change the state of joint system $AB$. This is mainly because, without a method to track information that was originally stored in $A$, the ostensible information exchange between $A$ and $B$ yields no detectable difference.

Especially, if we start from an initial state $\rho_{RA}\otimes\sigma_B$ where $R$ is a reference system of $A$ and apply a catalytic unitary $M_{AB}$, then the information source $B$ gets correlated with $RA$ in the tripartite state $\sigma_{RAB}:=(\id_R\otimes \Ad_M)(\rho_{RA}\otimes\sigma_B)$. Any unitary that utilizes the information of $B$ in $\sigma_{RAB}$ must be compatible with it on $B$, so, due to the following Corollary of Theorem \ref{thm:semthm}, the marginal state on $RB$ does not change after the second step; it stays in the product state $\sigma_{RB}=\sigma_R\otimes\sigma_B$, which means that no information in $A$ has been transferred to $B$.

\begin{corollary} \label{coro:symeq}
    If $\mds{1}_R\otimes U_{AB}$ with $U\in\mf{U}(AB)$ utilizes only semantic information of $B$ in $\sigma_{RAB}$, then we have
    \begin{equation}
        \Tr_A[\Ad_{U_{AB}} \circ \mcal{L}_A (\sigma_{RAB})]=\Tr_A[\mcal{L}_A(\sigma_{RAB})],
    \end{equation}
    for any $\mcal{L}\in\mf{L}(A)$. Especially, when $\mcal{L}=\id_A$, we get
    \begin{equation}
        \Tr_A[\Ad_{U_{AB}} (\sigma_{RAB})]=\sigma_{RB}.
    \end{equation}
\end{corollary}

Even after this observation, we should remark that Definitions (S:A-C) do not guarantee that there is no influx of information into the randomness source at all. Information that was encoded in the correlation between the source and the target system can be concentrated into the source.

For example, in the Szilard engine example we discussed, ((\ref{eqn:szc})-(\ref{eqn:sza})), if we call the purifying system of (\ref{eqn:szb}) $R$, then $I(R:C)$ increases from 1 bit to 2 bits in the course of interaction between $C$ and $G$, although we interpreted that no physical property other than information of $C$ was used in the interaction. This is not because information flowed from $G$ to $C$, but because the quantum entanglement of $CG$ with $R$ was concentrated into $C$ after the interaction, albeit it was not accompanied by information flow form $G$ to $C$.

We can interpret Definition (S:B) as that we characterize usage of (pure) semantic information of $B$ in $\sigma_{AB}$ as an interaction in which no information in $AB$ that is \textit{also present in} $A$ flows to $B$. Corollary \ref{coro:seminf} easily follows from Definition (S:B). Proof is given in Appendix. 

\begin{corollary}  \label{coro:seminf}
If a bipartite unitary operation $\mcal{U}=\Ad_U$ with $U\in\mf{U}(AB)$ utilizes \textit{(semantic) information} of $B$ in a bipartite state $\sigma_{AB}$, then, for any extension  $\sigma_{RAB}$ of $\sigma_{AB}$ such that $I(R:A)=I(R:AB)$, we have
\begin{equation} \label{eqn:seminf}
    \Tr_A[(\id_R\otimes\mcal{U})(\sigma_{RAB})]=\sigma_{RB}.
\end{equation}
\end{corollary}

As it was shortly discussed in Ref. \cite{lie2021correlational}, a randomness source correlated with a target system can absorb randomness as demonstrated in the example of Szilard engine initializing a gas molecule. This is impossible with uncorrelated randomness sources since they can only increase the amount of randomness in the target system. Now, with the complete characterization of information usage in correlated quantum system, we can quantify the amount of randomness that a given source can absorb or yield.

\begin{theorem} \label{thm:leasod}
    The least disordered state on $A$ that can be made from $\sigma_{AB}$ using $B$ as an information source is $\sum_j\left(\sum_i p_i \lambda_j (\sigma_A^i)\right) \dyad{j}_A$ where $\sigma_{AB}=\sum_i p_i \sigma_{AB}^i$ is the essential decomposition of $\sigma_{AB}$ on $B$. 
\end{theorem} 
Proof can be found in Appendix. Theorem \ref{thm:leasod} shows that quantum correlation is useless for catalytic randomness absorption. Only classical correlation between $A$ and $B$, which provides deterministic protocol to align eigenbases of conditional states of $A$, can reduce the amount of randomness in $A$ without leaking any information of it to $B$. Why is it so? Classical information can be copied and deleted, unlike quantum information, so reduction of randomness in $A$ can happen without any change in $B$ when it is conditioned on classical data in $B$.

 It is important that the results of this Section do not imply that pure entangled states allow no utilization of semantic information of any form whatsoever. We expect that there is a multitude of information flow in generic quantum interactions, but they are often too complicated and complex in both directions, or, sometimes, in ambiguous directions. Therefore, to understand the nature of (quantum) information flow, we only focused on directional information flow, which also has characterization as pure information usage. It is only that utilization of semantic information in pure multipartite states necessitates physical manipulation of information carrier.

We remark that our usage of the term \textit{semantic information} may not completely agree with others; we used the term to refer to information contained in a system that is correlated with another system the agent is going to interact with. This correlation differs from correlation among subsystems of a information source considered in delocalized catalysis of randomness. Our definition of semantic information is not propositional, hence cannot be true or false on its own. Hence, our semantic information does not satisfy the criteria of Floridi \cite{floridi2013philosophy}. One might think that our semantic information is closer to what Floridi calls \textit{environmental} information.

Nevertheless,  well-formedness can be expressed in terms of syntax, i.e. correlation between subsystems of information source like that between a sentence and the language, and semantic information given as multipartite state is meaningful as it is informative about the world outside of information source and as truthful as the given state describes the physical reality. This type of probabilistic and correlational definition was necessary for the generalization to quantum semantic information. In summary, our `semantic information' does not refer to the essence of information that is exclusively semantic but refers to information that \textit{could} contain semantic content.

\subsection{Superselection rules in delocalized and dynamical catalyses}
The essential decomposition for bipartite states already identifies the partition of the Hilbert spaces that should be essentially classically distinguishable, but there could be additional classical structure imposed by the superselection rule of each system. This consideration was made in identifying catalysis sector for static and local catalysis of randomness in Section \ref{subs:catran}. For delocalized catalysis of randomness, we modify Definition \ref{def:essdec} suitably.

\begin{definition} \label{def:supdef}
     For systems $A$ and $B$ in state $\rho_{AB}$ with the essential decomposition $A=\bigoplus_{i\in\mcal{I}_A\cup\mcal{II}_A} A_i$, suppose that there is a superselection rule with the superselection sectors $A=\bigoplus_j A'_j$. We let $A^\circ_{(i,j)}:= A_i\cap A'_j$ for $i\in\mcal{II}_A$ and all $j$, and let $\{(i,j)\}_{i\in\mcal{II}_A,j}$ be the new $\mcal{II}_A$. Then, the finer decomposition $A=\bigoplus_{i\in\mcal{I}_A} A_i \oplus \bigoplus_{k\in\mcal{II}_A} A^\circ_k$ is the essential decomposition under the superselection.
\end{definition}

Note that the superselection sectors cannot intersect nontrivially with type I subspaces of essential decompositions as the quantum state in each subspace cannot be a PC-Q state, hence no superselection rule can be nontrivially imposed on it. Physically, superselection rules only limit the quantum advantage that can be taken from type II subspaces by partitioning a large uniform quantum states into the tensor product of smaller ones and forbidding nonclassical interaction between them. Since the catalytic entropies of quantum channels are defined through the delocalized catalytic entropies of their corresponding Choi matrices, this new definition equally affects the definition of the dynamical catalytic entropies.

Definition \ref{def:supdef} provides a rather complicated way of treating randomness sources under superselection rules, but we show that actually it can be unified within the formalism of delocalized catalysis of randomness. When $\{Q_i\}$ are projectors onto superselection sectors of $A$, then any given catalysis $\rho_{AB}$ can be replaced with an extension $\rho_{E_A AB}$ given as
\begin{equation}
    \rho_{E_AAB}=\sum_i \dyad{i}_{E_A}\otimes (\Ad_{Q_i}\otimes\id_B)(\rho_{AB}),
\end{equation}
when it is treated as a delocalized randomness source. It can interpreted that the classical observable $i$ of $A$ which is forbidden to be in superposition should be treated as a piece of classical data correlated with the quantum state being used as a catalyst. Thus, introduction of delocalized catalysis of randomness nullifies the necessity of introducing $C^*$-algebra formalism to discuss about catalysts under superselection rules.

\subsection{The no-stealth theorem}

\begin{figure}[t]
    \includegraphics[width=.4\textwidth]{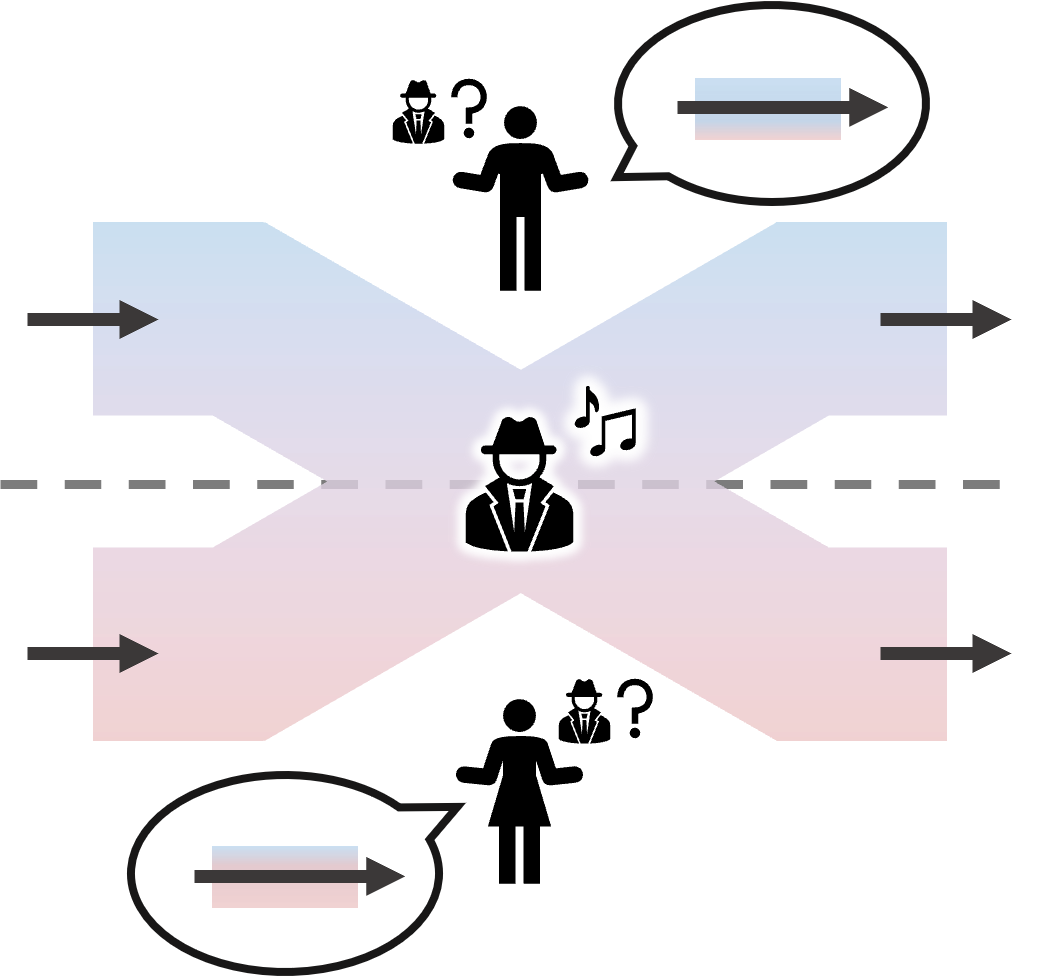}
    \caption{Suppose that input and output systems of a given quantum operation are reversibly distributed to two systems. Is it possible to hide the identity of the operation from the respective systems? In other words, is it possible to implement quantum operations stealthily? The no-stealth theorem says that it is impossible.}\label{fig:stealth}
\end{figure}

We consider the following dynamical generalization of the no-hiding theorem  \cite{braunstein2007quantum}, or equivalently, the no-masking  theorem\cite{modi2018masking}. Consider that we want to hide a dynamical process $\mcal{N}\in\subc(A)$ from two parties $A$ and $B$ by applying a global superunitary operation $\Omega\in\mf{SC}(AB)$. (Alternatively one could an arbitrary consider multipartite channel $\mcal{N}$. See Appendix \ref{subs:CPmap}.) By hiding, we mean that both of the marginal processes are constant regardless of the process $\mcal{N}$, (See FIG. \ref{fig:stealth}.) i.e.,
\begin{equation}
    \STr_B[\Omega(\mcal{N}_A\otimes\id_B)]=\STr[N]\mcal{E},
\end{equation}
and
\begin{equation}
    \STr_A[\Omega(\mcal{N}_A\otimes\id_B)]=\STr[N]\mcal{F},
\end{equation}
for some channels $\mcal{E}\in\mf{C}(A)$ and $\mcal{F}\in\mf{C}(B)$ and for all $\mcal{N}\in\subc(A)$. As discussed in Section \ref{subs:dyca}, the duality between delocalized and dynamical settings immediately yields that it is equivalent to the problem of hiding a bipartite state $\rho_{AA'}$, i.e., with some unitary operators $U_0\in\mf{U}(AB)$ and $U_1\in\mf{U}(A'B')$, we want
\begin{equation}
    \Tr_{BB'}[\Ad_{U_0\otimes U_1}(\rho_{AA'}\otimes \phi_{BB'}^+)]=\eta_{AA'},
\end{equation}
and
\begin{equation}
    \Tr_{AA'}[\Ad_{U_0\otimes U_1}(\rho_{AA'}\otimes \phi_{BB'}^+)]=\zeta_{BB'},
\end{equation}
for some quantum states $\eta_{AA'}$ and $\zeta_{BB'}$. This type of processes were called a randomness-utilizing processes in Ref.\cite{lie2020uniform}, and it was shown there that every dimension preserving randomness utilizing process must be a catalysis. Hence, we can set $\zeta_{BB'}=\phi_{BB'}^+$, which is a pure state. Also, because the delocalized catalytic entropy of $\phi_{BB'}^+$ is zero, $\eta_{AA'}$ cannot have larger entropy than the input state $\rho_{AA'}$, which can be chosen as a pure state, hence $\eta_{AA'}$ must be pure as well. This immediately yields a contradiction, since whenever $\rho_{AA'}$ is mixed, then the transformation $\rho_{AA'}\mapsto\eta_{AA'}$ decreases the entropy, which is impossible with a catalytic map. Remember that every catalytic map is unital, so it cannot decrease the entropy of the input state.

It follows that the original task of hiding arbitrary quantum process $\mcal{N}\in\subc(A)$ by unitarily distributing it to two parties is also impossible. In short, a quantum process cannot be stealthy on a system with reversible time evolution. Nevertheless, by using the resource theory of randomness for quantum processes developed in Section \ref{subs:dyca}, it is indeed possible to hide quantum processes when there is a randomness source with enough randomness.

\subsection{Examples} \label{sec:examples}
First, any pure state shared between two parties is useless as a randomness source. Especially, the maximally entangled state, corresponding to the identity channel through the Choi-\jami isomorphism, cannot yield any information without being perturbed.

On the contrary, every classical-classical (C-C) state can yield all of its entropy through catalysis. Suppose that a quantum state $\sigma_{AB}^{cc}$ is a C-C state:
\begin{equation}
    \sigma_{AB}^{cc}=\sum_{i,j} p(i,j)\dyad{i}\otimes\dyad{j},
\end{equation}
with the superselection rules that forbid any superposition between basis elements (i.e. $\{\ket{i}\})$ for both systems. For $\sigma_{AB}^{cc}$, every $\text{Span}{\{\ket{i}\}}$ for both systems is type II subspace with dimension 1, therefore the delocalized catalytic entropies and the ordinary entropies are the same, i.e., $\dent{\sigma_{AB}^{cc}}=\rent{\sigma_{AB}^{cc}}=\rent{\{p(i,j)\}_{i,j}}$ for all $\alpha\geq 0$.

This fact could be directly translated to classical-to-classical channels. Suppose that $\mf{B}(B)$ is the $C^*$-algebra of $|B|$-dimensional diagonal matrices and $\mcal{R}_c\in\mf{C}(B)$ is a classical channel;
\begin{equation}
    {\mcal{R}_c}(\rho)=\sum_{i,j=1}^{|B|}p(j|i)\bra{i}\rho\ket{i}\dyad{j},
\end{equation}
for some conditional probability distribution $p(j|i)$. Then its Choi matrix is a C-C state, i.e., $J_{BB'}^{\mcal{R}_c}=|B|^{-1}\sum_{i,j}p(j|i)\dyad{j}_B\otimes\dyad{i}_{B'}$ and $J_{BB'}^{\mcal{R}_c}$, thus $\cent{\mcal{R}_c}=\rent{J_{BB'}^{\mcal{R}_c}}=\rent{\{|B|^{-1}p(j|i)\}_{i,j}}$ for all $\alpha\geq0$.

Next, suppose that systems have coarser superselection rules compared to completely classical systems. Let $A=\bigoplus_i A_i$ and $B=\bigoplus_j B_j$ be the superselection sectors of two systems with $\Pi_i^A:=\mds{1}_{A_i}$ and $\Pi_j^B=\mds{1}_{B_j}$. Consider any classically correlated state of the following form
\begin{equation}
    \sigma_{AB}^{pc}=\sum_{i,j}p(i,j)\pi_{A_i}\otimes\pi_{B_j}.
\end{equation}
Then, the corresponding DREO is unitarily similar to
\begin{equation}
    \mf{d}\left(\sigma_{AB}^{pc}\right)\approx\sum_{i,j}p(i,j)\pi_{A_i}^{\otimes2}\otimes\pi_{B_j}^{\otimes2},
\end{equation}
hence we have $\dent{\sigma_{AB}^{pc}}=\cent{\sigma_{AB}^{pc}}$ and $S^{\diamond\diamond}(\sigma_{AB}^{pc})=S(\sigma_{AB}^{pc})+\sum_{i,j}p(i,j)\log_2(|A_i||B_j|)$ when $\alpha=1$. This means that there is no impediment from the constraints imposed by the delocalized setting when there is no type I subspaces in the essential decompositions.

The channel counterpart is the following type of measure-and-prepare channel from $A$ to $B$ with the superselection rules $A=\bigoplus_i A_i$ and $B=\bigoplus_j B_j$,
\begin{equation}\label{eqn:pcchan}
    {\mcal{R}_{mp}(\rho)=\sum_{i,j} p(j|i)\Tr[\Pi_{i}^A\rho]\;\pi_{{B}_j},}
\end{equation}
for any conditional probability distribution $p(j|i)$. The Choi matrix of this channel has the following spectral decomposition,
\begin{equation}\label{eqn:pcspec}
    J_{BA}^{\mcal{R}_{mp}}=\sum_{i,j} p(j|i) a_i \pi_{B_j}\otimes \pi_{A_i},
\end{equation}
where $a_i:=|A_i|/|A|$. A special case is the completely depolarizing channel with no superselection rules and trivial measurement, i.e.,
\begin{equation}
    \mcal{R}_{cp}(\rho)=\pi_B\Tr[\rho].
\end{equation}
The catalytic entropy of this channel, which functions as the completely randomizing quantum channel, is $\cent{\mcal{R}_{cp}}=2\log_2|A|+2\log_2|B|$. However, if both systems $A$ and $B$ are classical, then the same channel $\mcal{R}_{cp}$ now models ``dice rolling'', and the catalytic entropy becomes the half; $\cent{\mcal{R}_{cp}}=\log_2|A|+\log_2|B|$.

Conversely, let us consider the pinching channel n with respect to a complete set of orthonormal projectors $\{\Pi_i\}$ on $B$ such that $\sum_i \Pi_i = \id_B$, i.e.,
\begin{equation}
    \mcal{R}_d(\rho)=\sum_i \Pi_i \rho \Pi_i.
\end{equation}
In this case, the Choi matrix is of the randomness source is 
\begin{equation}
    J_{BB'}^{\mcal{R}_d}=\sum_i b_i\dyad{\Gamma_i},
\end{equation}
 where $b_i:={|B_i|}/{|B|}$, $\ket{\Gamma_i}=|B_i|^{-1/2}(B_i\otimes\mds{1}_{B_i'})\sum_j\ket{jj}_{BB'}$ with $B_i=\supp{\Pi_i}$. Here, every subspace $B_i$ is either a type I or 1-dimensional type II subspace. Hence, $\cent{\mcal{R}_d}=\dent{J_{BB'}^{\mcal{R}_d}}=\rent{\{b_i\}}\leq\cent{J_{BB'}^{\mcal{R}_d}}$ for all $\alpha\geq0$. It means that even if there are multiple $b_i$ with the same value, i.e., even if $J_{BB'}^{\mcal{R}_d}$ has degeneracy, the quantum correlation between two systems hinders the utilization of that correlation without leaving traces.


\section{Discussion} \label{sec:disc}
\subsection{Physicality of information} \label{subsec:physical}
In the seminal article `Information is Physical' (1991) \cite{landauer1991information}, Landauer argued that information is physical by reciting the observations that there is no nontrivial minimal energy dissipation accompanying information processing tasks such as computation, copying and communication. These evidences imply that deletion of information is the only source of nontrivial energy cost, which supports the view that a certain amount of energy corresponds to a certain amount of energy, independent of how it is processed, in favor of the interpretation that information is a physical entity as matter is equivalent to energy through the mass-energy equivalence.

Certainly, Landauer's argument irrefutably shows that the \textit{presence} of information in our physical universe is necessarily physical as Landauer said ``Information is not an abstract entity but exists only through a physical representation'' \cite{landauer1999information}. However, the problem with this almost tautological usage of the term `physical' is that it makes every physically perceivable abstract concept physical. For example, \textit{money} can only exist through physical notes and coins or digitalized currencies in physical computers, and \textit{law} must be recorded on some physical representation and can only be enforced with physical methods by a \textit{government}, which is also an abstract concept that exists only through a physical manifestation. We can even say that every abstract concept that involves information exchanges is physical if information is physical. If every concept relevant to a physical agent counts as physical, then this notion of physicality might not be very useful as there would be virtually no nonphysical concept.

A more operational criterion for the physicality of concepts would be asking if usage or action with/involving the concept requires detectable change to physical representation of the concept that is unavoidable, even in an approximate sense. Perhaps, the term \textit{material} might be more appropriate to describe such a property since there are concepts of physical nature that are not material by themselves. For example, `solidness' is represented by a hammer used to drive a nail into the wall, but the hammer, in the practical sense, is not detectably altered after the process. Clearly, `solidness' is a property of physical nature but not a matter-like concept; `solidness' did not depart from the hammer to the wall like a particle. Likewise, every catalyst in chemistry and quantum resource theory is also not a physical representation of material concept, albeit they might play a physical role in the respective catalysis process. As a matter of fact, since the terms `physicalism' and `materialism' are often used interchangeably \cite{stoljar2001physicalism}, we will not introduce another term and call the property simply `physicality'. This is the perspective we take in this work about information, and the argument of Landauer ironically supports the claim that information is not physical in our sense, as Landauer argued that energy cost of information processing other than deletion can be made arbitrarily small. 

Our notion of physicality could be relative, as what is expected from an operational concept. Naturally, physicality of information now depends not only on the information storage but also on the method of utilizing it. For example, software, in contrast to hardware, is usually considered nonphysical because installation, execution and deletion of software leave no apparent physical trace on the hardware it is running on. However, of course, it is true not only that software accompanies physical traces on hardware detectable with careful inspection, but also one can physically interact with software through input and output devices, hence software is as physical as hardware for its user equipped with proper devices. 

We defined information as something that can spread from its source without altering it, hence it is required to be nonphysical by definition. Is this notion of information also relative? We first examine it for classical information. Let us consider the classical version of catalytic randomness. Consider interaction between system 1 and 2, where $(i,j)$ represents the situation where system 1 is in the state $i$ and system 2 is in the state $j$. We want to formulate a classical version of (\ref{eqn:catal1}) and (\ref{eqn:catal2}). Invertible classical operation is permutation, thus we let $f:(i,j)\mapsto (f_1(i,j),f_2(i,j))$ be a permutation of states of the joint system of 1 and 2, where system 1 is a target system and system 2 is a catalyst. When the initial probability distribution of system 2 is $(p_j)$, then the condition for $f$ to be catalytic permutation compatible with $(p_j)$ is
\begin{equation}
    \sum_{j':j=f_2(i,j')} p_{j'}=p_{j},
\end{equation}
for all $i$ and $j$. Similarly to catalytic quantum randomness, $f_2(i,\,\cdot\,)$ must preserve every non-degenerate probability distribution, and can permutate every degeneracy block of $(p_j)$ (the set of $j$ with the same probability $p_j$). As a special case, for the completely uniform distribution, $\pi_2$, every permutation $f$ such that $f_2(i,\,\cdot\,)$ is a permutation for every $i$ is catalytic permutation compatible with $\pi_2$. This fact may come off as weird to some readers, because permuting the outcomes of an information source may seem to leak information to the source. However, if the source is not correlated with any other information sources you have, then there is no way to tell if the permutation has taken place: You cannot tell if someone flipped the unknown outcome of a random coin toss.

Even if permutation of degenerate states of catalyst is allowed in pure information utilization, some readers might still wonder why would one want to do that. Indeed, reading a message and scrambling the letters of the message sound weird and look unnecessary when the purpose is simply extracting as much information as possible. In generic cases, however, this permutation is accidental rather than intentional. One can consider each state in each degeneracy block a microscopic state and each degeneracy block of $(p_i)$ a macroscopic state. Turning a page of a book will disturb the molecules in the paper even when it is done extremely carefully. But, if it can be done in a macroscopically undetectable fashion, then the action only permutes the microscopic states belonging to a same macroscopic state. Thus, it still counts as pure information utilization on this macroscopicity level.

The intuition that the permutation is invasive is not wrong nonetheless, as manipulating a part of a correlated information source can indeed leak information. If you tossed a coin and wrote down the outcome on a piece of paper, then then the coin and the paper are correlated. In this case, if someone flips the coin, then you can detect it by referring to the paper. Actually, this is exactly how classical secret sharing works; encoding information into correlation and correlation only. Nevertheless, if you cannot access the paper, then interactions that might flip the coin can still count as pure information utilization. This shows that physicality of classical information is also relative, because the choice of the system that you will treat as information source affects the physicality. 

Nonetheless, a question on the possibility of universally nonphysical classical information still remains: Is it possible to utilize information of a classical system regardless of its relation with the outer world? Indeed, every permutation $f$ that fixes every $j$, i.e. $f_2(i,j)=j$ for all $i$ is compatible with every extension of system 2, i.e. a combination of system 2 and any system 3 that is arbitrarily correlated with system 2. Such a permutation corresponds to simply `reading' $j$ and implementing a permutation on system 1 conditioned on $j$. One can easily see that this action never changes the joint probability distribution of system 2 and 3. This is the notion of classical information we are familiar with: information that can be freely read and distributed and does not necessitate a nontrivial minimum amount of physical effect on information carriers.

Does the same conclusion hold for quantum information? In our definition (see (\ref{eqn:catal2})), utilizing only information in quantum state $\sigma_B$ means leaking no information to it. In other words, we defined utilization of quantum information to be nonphysical as well. However, just like classical information sources, a quantum information source could be correlated with other systems, i.e., $\sigma_B$ could be a marginal state of its extension $\sigma_{AB}$. We can easily observe that interacting with a part of correlated information source exactly corresponds to delocalized catalysis of randomness and Theorem \ref{thm:delodep} says that TQ-TQ bipartite states cannot yield randomness through delocalized catalysis. But, since every mixed state $\sigma_{B}$ has a TQ-TQ extension $\sigma_{AB}$, namely, its purification. Hence, every utilization of quantum information can be detected by someone with enough amount of side information; there is no universally nonphysical quantum information, contrary to classical information. This observation can be summarized as follows.

\begin{theorem}
    For any catalysis unitary $U_0\in\mf{U}(A_0B_0)$ compatible with $\sigma_{B_0}$, there exists an extension $\sigma_{B_0B_1}$ of $\sigma_{B_0}$ such that $(U_0,U_1)$ is not compatible with $\sigma_{B_0B_1}$ for any $U_1\in\mf{U}(A_1B_1)$.
\end{theorem}

One of the goals of establishing the framework of catalysis of quantum randomness is to distinguish `quantum state' and `quantum information', two terms that are often mixed up in quantum information community. This distinction is needed since quantum state describes every physically accessible properties of a quantum system, be it informational or not.  Thus, accepting this distinction, the no-cloning theorem only forbids cloning of quantum state, not quantum information. In fact, the task of `cloning quantum information' must be carefully redefined. Nonetheless, the fact that there is no universally nonphysical quantum information hints that the gist of the no-cloning theorem still lives on for quantum information. The fact that cloning and distribution of classical state can be freely done strongly suggests that classical information is a nonphysical entity operationally independent of its physical representation, and vice versa. In contrast to this, quantum information is firmly bound to its physical representation, which can be interpreted to be strongly related to the fact quantum state is unclonable.

We may summarize the results of this Section with a slogan `quantum information is physical from a broader perspective' to emphasize the difference between classical and quantum information. In our formalism, pure information utilization is required to be nonphysical for a given information source in the first place, hence the slogan should be interpreted as that for every pure quantum information utilization there exists an agent who perceives it not as a purely informational interaction, whereas the same may not hold for classical information. After all, as we pointed out, physicality of information depends on its definition and perspective of user.

\subsection{Concave resource theories} \label{subsec:concave}

As it was briefly outlined in Introduction, we define \textit{concave resource theory} as a theory that consists of the state of \textit{resourceful states} $\mf{R}$ (``the resourceful set'') and the set of \textit{resourceful operations}, operations that preserve $\mf{R}$, $O_\mf{R}$. Here, the resourceful set $\mf{R}$ is required to be convex, i.e., if $\rho,\sigma\in\mf{R}$, then $\lambda\rho+(1-\lambda)\sigma\in\mf{R}$ for any $0\leq \lambda \leq 1$. Any state that is not resourceful is called \textit{free}. In contrast to the fact that usually the distance to the concave set of free states is used as a measure of resource, it is natural to measure how deep inside a state is placed in the resourceful set in a concave resource theory. The most typical concave resource theory would be that of entropies, whose resource measures are Schur-concave entropic quantities.


As entropic measures like the von Neumann entropy are already a well-studied topic, one might consider concave resource theories are more or less trivial. However, there could be still other types of resource theories of randomness and the theory of catalytic quantum randomness is one of it. Albeit it is a concave resource theory, catalytic entropies are not concave functions. For example, slightly mixing the maximally mixed state with a non-degenerate state significantly decreases its catalytic entropy because it destroys the degeneracy of it.


Nevertheless, we could anticipate that superunitary operations must be a part of free operations of generic concave resource theories. Our definition of superunitary operation does not have one of the most distinct characteristics of the physical implementation of superchannels: the effect of memory system. This is because discarding subsystem is no longer a free operation in resource theory of randomness. There is only one exception and that is discarding a quantum system that is not allowed to change it marginal state, because discarding such a system will not lead to any leakage of information, and that fits our definition of utilizing randomness and randomness only. (See Section \ref{subs:rai}.)

The resource theory of randomness (RTR), as a concave resource theory, has many implications that go against our intuition built from conventional convex resource theories. The resource in the RTR is randomness, which is not inherently a quantum property, hence not every object with large quantumness is superior compared to its classical counterpart. For example, a maximally entangled state shared by two parties, which is a very useful resource in entanglement theory, is completely useless in delocalized catalysis of randomness. In general, whenever there is quantum correlation in a bipartite quantum state, there exists a type I subspace in the essential decomposition, and it hinders catalytic extraction of randomness (See Section \ref{sec:examples}). It is because states with quantum correlation are sensitive to the action of local unital channels \cite{lie2021faithful}.

One should not understand it as that every quantumness is an obstacle in randomness extraction. For example, local coherence is helpful for maximizing extractable randomness of type II subspaces. This is the very reason why there are dimension-doubling effects in REO or DREO of randomness sources. However, again, it does not mean that coherence is already present in the state helps catalysis of quantum randomness. When we say that local coherence boosts catalytic quantum randomness, it means that exploiting coherent quantum operation boosts the efficiency of catalytic randomness extraction. The ambivalent roles of quantumness as presented here motivates the further study of quantum randomness to reveal its true nature and the extent of its power.

\subsection{Randomness amplification}
Suppose that there is a sequence of (classical or quantum) systems $(A_n)_{n=0}^\infty$, and the initial system is prepared in some state $\rho_0$. At step $n$, similarly with a Markov chain, only two adjacent systems $A_n$ and $A_{n+1}$ can unitarily interact with the constraint that information must not flow from $A_{n+1}$ to $A_n$. This means that catalysis of randomness should happen with system $A_n$ being the catalyst. Let $\rho_n$ be the state of $A_n$ after the interaction with $A_{n-1}$. We will call this type of sequence a \textit{randomness chain}.

Assume that $A_0$ is the only initial randomness source, i.e, every $A_n$ with $n\geq 1$ is prepared in a pure state. One observation we can make is that, when every system $A_n$ is classical, the amount of randomness never increases with increasing $n$. This is because $\cent{\rho_n}=\rent{\rho_n}$ for classical systems but $\rent{\rho_{n+1}}\leq\cent{\rho_n}$ by Theorem \ref{thm:catent}. On the other hand, if every system $A_n$ is quantum, then the amount of randomness can increase \textit{exponentially} over $n$. This is because $\cent{\rho_n}\geq\rent{\rho_n}$ and even $\cent{\rho_n}=2\rent{\rho_n}$ is achievable. In other words, \textit{randomness amplification} is possible only in the chain of quantum systems.

Interpretations of this observation could vary. One could conclude that in classical chain, when information back-flow is not allowed, then the total amount of information measured by its randomness can only decay over successive transmission between systems. It is fundamentally because classical systems cannot generate new randomness without shifting information to other systems. However, in quantum systems, correlation can be formed within a single system without requiring any randomness, in contrast to classical systems. Therefore, by using preexisting randomness, one can destroy the correlation and create even larger randomness. As a result, quantum randomness that was initially minuscule can be amplified to the macroscopic randomness after the long chain of quantum systems, but no information has flowed backward through the chain. 

Because of the generalization developed in this work, we can see that the same phenomenon could also to a chain of quantum processes. Analogously we can consider a sequence of quantum channels $(\mcal{N}_n)_{n=0}^\infty$ where $\mcal{N}_n\in\mf{C}(A_n)$ and there exists a catalytic superchannel $\Theta_n$ such that $\Theta_n(\mcal{M})=\STr[\Omega_n(\mcal{M}\otimes\mcal{N}_n)]$ with some catalysis superunitary operation $\Omega_n\in\mf{SL}(A_{n+1}A_n)$ compatible with the catalyst $\mcal{N}_n$ for every $n\geq 0$ so that $\Theta_n(\Upsilon_n)=\mcal{N}_{n+1}$ for some superunitary operation $\Upsilon_n\in\mf{U}(A_n)$. It means that all the randomness of $\mcal{N}_{n+1}$ is catalytically extracted from $\mcal{N}_n$, hence there is no detectable effect left on the action of $\mcal{N}_n$ alone by the randomness extraction. We will call this a randomness chain of quantum channels. For example, a depolarizing noise on a 1000-qubit quantum system can be realized from a depolarising noise on a qubit system after about 10 steps along a randomness chain because of the exponential growth of randomness. Along with chaos, this type of quantum randomness amplification might be one of the mechanisms realizing macroscopic disorder with microscopic initial disorder. An interesting observation is that a chain of completely dephasing channels cannot see this kind of randomness amplification because there are no type II subspaces that could yield quantum advantage of randomness extraction (See Section \ref{sec:examples}).

\section{Conclusions} \label{sec:conc}
Why is it important to understand what it means to use information and information only? With the success of quantum information theory, there has been a trend of calling the advantage of using quantum systems compared to using classical systems for implementing the same task the advantage of `quantum information', even when it is accompanied by destruction or deterioration of quantum systems. But after a moment's thought, not every quantum property is purely informational, and there is a necessity of distinguishing the power of information and that of other physical properties. In this work, following the gist of Shannon \cite{shannon1948mathematical}, we analyzed randomness as information in the quantum setting.

We generalized the resource theory of catalytic quantum randomness to delocalized and dynamical randomness sources. The delocalized and dynamical catalytic entropies were introduced to measure the catalytically extractable randomness within multipartite quantum states. In contrast to static catalysis of randomness, not every mixed state can yield catalytic randomness in the delocalized setting for nonclassically correlated quantum states are sensitive to the effect of catalytic maps. As an application, we proved a no-go theorem that is a generalization of the no-hiding theorem \cite{braunstein2007quantum}, the no-stealth theorem, that forbids unitarily hiding quantum processes by distributing it to two delocalized parties.

Furthermore, we critically examined the slogan `information is physical' by Landauer \cite{landauer1991information, landauer1999information}, and concluded that when we focus more on utilization of information rather than on mere presence of information, classical information is rather nonphysical, or can be made nonphysical with proper optimization, independently of perspective. On the other hand, we showed that utilization of quantum information cannot be universally deemed to be nonphysical. It is essentially because quantum correlation, especially entanglement of pure quantum state, is strong enough to remember every action acted on a local system.

We also attempted to analyze semantic information in the context of catalytic randomness, by focusing on the correspondence between information's meaning and correlation with other systems. By doing so, we showed that non-semantic information, randomness, is a special case of semantic information and revealed that the usability of semantic information is exactly same with that of non-semantic information.

Models of information used in this work are rather too simple to cover every aspect of information theory and one may find the definition of information given in this work unsatisfactory or even disagree with it. Nonetheless, we reckon that the framework developed here successfully captures a certain aspect of information as a relatively nonphysical entity whose physical representation can affect the physical world without being affected and is presented in a concise modern quantum information language easily accessible by physicists. Indeed, the field of information theory is so vast that a single definition of information cannot explain every aspect of information as Shannon warned:
\begin{center}
    \textit{``It is hardly to be expected that a single concept of information would satisfactorily account for the numerous possible applications of this general field.''}
    
    --- Shannon (1953) \cite{shannon1953lattice}.
\end{center}

As we have completed the characterization of maximum entropy extractable with exact catalysis, natural next steps include generalization to approximate catalysis and the converse problem. By converse problem, we mean characterizing randomness sources that can realize a given catalytic map. 

Characterizing tasks that can be done without altering randomness sources is important for understanding the nature of randomness in physics in comparison to other concepts, but in practice, one can always use randomness in combination with other physical properties, hence it would be interesting to study the relation of the randomness cost and other costs of implementing quantum processes.

\begin{acknowledgements}
This work is largely based on the doctoral thesis of SHL. SHL thanks Seongwook Shin for helpful discussions. This work was supported by National Research Foundation of Korea grants funded by the Korea government (Grants No. 2019M3E4A1080074, No. 2020R1A2C1008609 and No. 2020K2A9A1A06102946) via the Institute of Applied Physics at Seoul National University and by Ministry of Science and ICT, Korea, under the ITRC (Information Technology Research Center) support program (IITP-2020-0-01606) supervised by the IITP (Institute of Information \& Communications Technology Planning \& Evaluation) This work is also supported by the quantum computing technologydevelopment program of the National Research Foundation of Korea(NRF)funded by the Korean government (Ministry of Science and ICT(MSIT)) (No.2021M3H3A103657312).

\end{acknowledgements}

\appendix

\section{Mathematical results}

\subsection{Issues of CP map input} \label{subs:CPmap}
In contrast to static catalysis, which requires the invariance of the state of randomness source for every normalized input state, we required dynamical catalysis the invariance of the randomness source channel for every CP trace nonincreasing map in (\ref{eqn:dycat2}). However, in contrast to that every subnormalized quantum state can be made into a normalized one by simply multiplying by some positive number, not every CP map can be made into a quantum channel (CPTP map) in the same way. Hence, one might suspect that requiring condition (\ref{eqn:dycat2}) for every $\mcal{N}\in\subc(A)$ is too severe. In this Section, we justify this condition. Alternatively, we could require the following condition,
\begin{equation} \label{eqn:cpr1}
    \STr_A\left[(\mf{id}_{E_0\to E_1}\otimes\Omega)(\mcal{N} \otimes \mcal{R})\right]= \STr_A[\mcal{N}]\otimes\mcal{R}
\end{equation}
for every $\mcal{N}\in\mf{C}(AE_0,AE_1)$, where $\mf{id}_{E_0\to E_1}(\mcal{L})=\id_{E_1}\circ\mcal{L}\circ\id_{E_0}$ for every $\mcal{L}\in\mf{L}(E_0,E_1)$. The differences are that now $\mcal{N}$ is a multipartite channel, and that output channels $\STr_A[\mcal{N}]\in\mf{C}(E_0,E_1)$ and $\mcal{R}\in\mf{C}(B)$ are required to be uncorrelated.

This is a well-motivated requirement, since superchannels can be applied to a part of multipartite channels, and the requirement of information non-leakage through $\Omega$ can be re-interpreted as the requirement of no formation of correlation between the systems that did not interact directly through $\Omega\in\mf{SC}(AB)$. We remark that any CP trace nonincreasing map can be a subchannel of another channel. It means that for any $\mcal{N}_0\in\subc(A)$, there exists some $\mcal{N}_1\in\subc(A)$ such that $\mcal{N}_0+\mcal{N}_1\in\mf{C}(A)$. Also, for some $U\in\mf{U}(AE_0)$ and a POVM $\{M_0,M_1\}$ with $M_0+M_1=\mds{1}_{E_0}$ on $E_0$ and
\begin{equation}
    \mcal{N}_i(\rho) = \Tr_{E_0}[(\mds{1}_A\otimes M_i)\Ad_U(\rho\otimes\dyad{0}_{E_0})],
\end{equation}
for every $\rho\in\mf{B}(A)$ and $i=0,1$. Naturally, we can define the corresponding channel $\mcal{N}\in\mf{C}(A,AE_1)$ given as
\begin{equation}
    \mcal{N}(\rho):=\mcal{N}_0\otimes\dyad{0}_{E_1}+\mcal{N}_1\otimes\dyad{1}_{E_1}.
\end{equation}
With this expression, (\ref{eqn:cpr1}) requires that
\begin{equation} \label{eqn:both}
    \STr_A\left[(\mf{id}_{E_0\to E_1}\otimes\Omega)(\mcal{N} \otimes \mcal{R})\right]= \sigma_{E_1}\otimes\mcal{R},
\end{equation}
with $\sigma_{E_1}=\STr[\mcal{N}_0]\dyad{0}_{E_1}+\STr[\mcal{N}_1]\dyad{1}_{E_1}$. However, we can observe that
\begin{equation}
    \bra{i}_{E_1}\mcal{N}\ket{i}_{E_1}=\mcal{N}_i,
\end{equation}
for $i=0,1$, therefore by contracting $\dyad{i}_{E_1}$ with the both sides of (\ref{eqn:both}), using $\bra{i}_{E_1}\sigma_{E_1}\ket{i}_{E_1}=\STr[\mcal{N}_i]$, we get
\begin{equation}
    \mf{Tr}_A\Omega(\mcal{N}_i \otimes \mcal{R})= \STr[\mcal{N}_i]\mcal{R}.
\end{equation}
Since $\mcal{N}_0$ was chosen arbitrarily in $\subc(A)$, we can see that (\ref{eqn:cpr1}) implies condition (\ref{eqn:dycat2}).

Conversely, let $\mcal{L}^\ddag:=\dag\circ\mcal{L}\circ\dag$ for any linear map $\mcal{L}$. We can see that any linear map $\mcal{L}$ can be decomposed into the Hermitian-preserving part $\mcal{L}_R:=(\mcal{L}+\mcal{L}^\ddag)/2$ and the anti Hermitian-preserving part $\mcal{L}_I:=-i(\mcal{L}-\mcal{L}^\ddag)/2$ so that $\mcal{L}=\mcal{L}_R+i\mcal{L}_I$. Again, any Hermitian-preserving linear map $\mcal{H}$ can be expressed as the difference of two CP maps $\mcal{P}$ and $\mcal{L}$ so that $\mcal{H}=\mcal{P}-\mcal{N}$. (It follows from the spectral decomposition of its Choi matrix.) Hence, if (\ref{eqn:dycat2}) holds for every $\mcal{N}\in\subc(A)$, by the linearity, it also holds for every $\mcal{L}\in\mf{L}(A)$, so (\ref{eqn:cpr1}) follows. Therefore, (\ref{eqn:dycat2}) and (\ref{eqn:cpr1}) are equivalent.

\subsection{Proof of Proposition \ref{prop:nogen}}
\begin{proof}
    Let $\mcal{C}\in\mf{C}(X)$ be a catalytic map. The entropy increase of a quantum state $\sigma_X$ by $\mcal{N}$ cannot be larger than that of its purification $\ket{\Sigma}_{XX'}$ $(\Tr_{X'}\dyad{\Sigma}_{XX'}=\sigma_X)$ \cite{lie2021correlational}. Therefore, the largest entropy production happens on a pure bipartite state, and let $\ket{\Psi}_{XX'}$ be a pure state that achieves the maximum entropy production by $\mcal{N}$. Note that every pure bipartite state s related with a maximally entangled state $\ket{\Phi}_{XX'}$ by the action of a local matrix, i.e. there exists $M\in\mf{B}(X)$ such that $\ket{\Psi}_{XX'}=(\mds{1}_X\otimes M_{X'})\ket{\Phi}_{XX'}$. Note that $\mcal{N}$ cannot generate any randomness if $\mcal{N}_X(\dyad{\Psi})_{XX'}$ is pure, i.e., rank 1. Since $\mcal{N}_X(\dyad{\Psi}_{XX'})=(\id_X \otimes \Ad_{M})(\mcal{N}_X(\dyad{\Phi}_{XX'}))$, if $\mcal{N}_X(\dyad{\Phi}_{XX'})$ is pure, then it follows that $\mcal{N}$ cannot generate randomness. Conversely, if $\mcal{N}$ cannot generate randomness, then by definition $\mcal{N}(\dyad{\Phi}_{XX'})$ is pure.
\end{proof}

\subsection{Discssion on M{\o}lmer's conjecture}
 M{\o}lmer's conjecture \cite{molmer1997optical} insists that the quantum state of laser light should not be represented by a pure coherent state
\begin{equation} \label{eqn:alpha}
\ket{\alpha} = e^{-\frac{|\alpha|^2}{2} }{\sum_{n=0}^{\infty}\frac{\alpha^n}{\sqrt{n!}}\ket{n}},
\end{equation}
but the mixed state
\begin{equation}
    \frac{1}{2\pi}\int_{0}^{2\pi}\ket{|\alpha|e^{i\theta}}\bra{|\alpha|e^{i\theta}}d\theta=e^{-|\alpha|^2}\sum_{n=0}^{\infty}\frac{|\alpha|^{2n}}{n!}\ket{n}\bra{n},
\end{equation}
because of the loss of phase information caused by inaccessibility of laser device. Choosing to use the pure coherent state representation without considering correlated systems amounts to committing the \textit{preferred ensemble fallacy} \cite{kok2000postselected, nemoto2004quantum}. 
When it is stated that `a random pure state $\ket{\phi}_A$ is prepared', oftentimes it is assumed, very implicitly, that there exists a fixed preparation protocol that produces $\ket{\phi}_A$. This protocol can be classically identified with a careful inspection, and be represented by an orthonormal basis $\{\ket{``\phi"}_P\}$ that is orthogonal between each different state, even when $\ket{\phi}_A$ itself is not orthogonal to each other, i.e., $\bra{``\phi"}\ket{``\psi"
}=0$ whenever $\phi\neq\psi$. In this case, the global quantum state of system $AC$ is
\begin{equation}
    \sum_\phi p(\phi) \dyad{\phi}_A\otimes \dyad{``\phi"}_C,
\end{equation}
with some probability distribution $p(\phi)$. (One can replace the sum with an integral when the probability distribution is not discrete.) If one runs the same preparation protocol $n$ times, then it becomes
\begin{equation}
    \sum_\phi p(\phi) \dyad{\phi}_A^{\otimes n}\otimes \dyad{``\phi"}_C.
\end{equation}
Or, systems $AC$ can even be entangled;
\begin{equation}
    \sum_\phi \sqrt{p(\phi)} \ket{\phi}_A\otimes \ket{``\phi"}_C.
\end{equation}
Whether to treat the whole system $AC$ or system $A$ alone as the information source depends on one's choice and on a given situation. For example, if it is implicitly assumed that there exists a referee who remembers the identity of the random state $\ket{\phi}_A$ and if you treat the relation between the state and the referee as a part of information you utilize, then the whole system $AC$ should be considered an information source. However, if system $A$ is in isolation from any context other than the distribution $p(\phi)$, then it is natural to treat only system $A$ as an information source.

\subsection{Proof of Theorem \ref{thm:delodep}}

\begin{proof}
The assumption that no randomness can be catalytically extracted from $\sigma_{B_0B_1}$ means that any catalytic unitary operators compatible with $\sigma_{B_0B_1}$ is a product unitary operator. Therefore, any action applied to $\sigma_{B_0B_1}$ is also of the form of product unitary operations, i.e. $\sigma_{B_0B_1}\mapsto \Ad_{V_0\otimes W_1}(\sigma_{B_0B_1})$ with some $V_0\in\mf{U}(B_0)$ and $W_1\in\mf{U}(B_1)$. As a special case, assume that $U_0=\mds{1}_{A_0B_0}$. It implies that $V_0=\mds{1}_{B_0}$. Note that, in this case, $W_1$ should be also proportional to the identity operator. It is because, if $W_1\not\propto \mds{1}_{B_1}$, then the random unitary operation given as $\frac{1}{2}(\id_{B_1}+\Ad_{W_1})$ on $B_1$  that is not a unitary operation also fixes $\sigma_{B_0B_1}$. This contradicts the previous result that any action on $\sigma_{B_0B_1}$ should be a unitary operation. It is equivalent to saying that whatever catalytic map is applied to system $B_1$, if it fixes $\sigma_{B_0B_1}$, then it should the identity operation. This property is called sensitivity to catalytic maps according to the definition given in Ref. \cite{lie2021faithful}. As the set of catalytic map is contained in the set of unital maps, and contains the set of random unitary operations, by the results of Ref.\cite{lie2021faithful}, it follows that it is equivalent to that $\sigma_{B_0B_1}$ is not a Q-PC state. The same argument can be applied when the roles of $B_0$ and $B_1$ are switched, thus $\sigma_{B_0B_1}$ is neither a PC-Q state.

Conversely, assume that $\sigma_{B_0B_1}$ is TQ-TQ. Let $U_0\in\mf{U}(A_0B_0)$ and $U_1\in\mf{U}(A_1B_1)$ be arbitrary catalytic unitary operators and  $\mcal{N}_0:=\mf{Tr}_{A_0}\circ\Ad_{U_0}\in\mf{UC}(B_0)$ and $\mcal{N}_1:=\mf{Tr}_{A_1}\circ\Ad_{U_1}\in\mf{UC}(B_1)$ be induced catalytic maps on $B_0$ and $B_1$ respectively. For $\sigma_{B_0B_1}$ to be compatible with $U_0$ and $U_1$, $\mcal{N}_0\otimes\mcal{N}_1$ must fix $\sigma_{B_0B_1}$. However, since catalytic maps can never decrease the von Neumann entropy, it means that both $\mcal{N}_0$ and $\mcal{N}_1$ fix the von Neumann entropy of $\sigma_{B_0B_1}$. By Theorem 2.1 of Ref.\cite{zhang2011neumann}, it is equivalent to that both $\mcal{N}_0^\dag\circ\mcal{N}_0$ and $\mcal{N}_1^\dag\circ\mcal{N}_1$ fix $\sigma_{B_0B_1}$. Since $\sigma_{B_0B_1}$ is TQ-TQ, it is sensitive to unital channels on both sides \cite{lie2021faithful}, hence it follows that  $\mcal{N}_0^\dag\circ\mcal{N}_0=\id_{B_0}$ and $\mcal{N}_1^\dag\circ\mcal{N}_1=\id_{B_1}$. It is equivalent to that both $\mcal{N}_0$ and $\mcal{N}_1$ are unitary operations, therefore $U_0$ and $U_1$ are product unitary operators. It follows that no randomness can be extracted from $\sigma_{B_0B_1}$.
\end{proof}

\subsection{Uniqueness of essential decomposition} \label{app:unique}

Recall the three criteria, $(i),$ $(ii)$ and $(iii)$ of Definition \ref{def:CPQ}. Because of $(i)$ and $(iii)$, any two essential decompositions of the same state should commute with each other. (By saying decompositions commute with each other, we mean that projectors corresponding to their subspaces are mutually commutative.) If their type I subspaces do not match, then, since $\rho_{AB}$ is still generalized block-diagonal with the intersections of both decompositions, some of type I component of $\rho_{AB}$ permits further decomposition on $A$, hence become a PC-Q state, which violates $(ii)$.

If there is a mismatch of type II subspaces between two essential decompositions (However, their spans should match because they are the perpendicular complement of the span of the same type I subspaces because of the previous paragraph), it leads to a violation of $(iii)$ since there always is a projector that commutes with one decomposition that does not with the other. Say, $A_1$ is a type II subspace of the first decomposition that intersects with two type II subspaces of the second decomposition, $A'_1$ and $A'_2$. Pick arbitrary $\ket{\phi_1}\in A_1\cap A'_1$ and $\ket{\phi_2} \in A_1 \cap A'_2$, and let $P$ be the rank-1 projector onto $\ket{\phi'}:=2^{-1/2}(\ket{\phi_1}+\ket{\phi_2})$. Since $\ket{\phi'}$ lies in $A_1$ (on which $\rho_{AB}$ is proportional to the identity operator on $A$ so that it commutes with every operator on $A_1$), $P$ commutes with $\rho_{AB}$ but it does not commute with the projectors onto $A'_1$ or $A'_2$. Thus, the essential decomposition is unique.

\subsection{Other results on essential decomposition} \label{app:ess}
The following two Propositions are not directly used in the rest of Section, but give insight into the structure of bipartite fixed points of quantum channels.
\begin{prop}
    If the product of unital maps $\mcal{N}\otimes\mcal{M}$ fixes a quantum state $\rho_{AB}$, then they also fix every projector onto each of its eigenspaces, $\Pi_i$. Also, $\mcal{N}$ fixes $\Tr_B\Pi_i$ and $\mcal{M}$ fixes $\Tr_A\Pi_i$. 
\end{prop}
\begin{prop}
    Assume that $\Pi$ is a projector on $AB$ such that $\Tr_B \Pi \propto \mds{1}_A$ and $\Tr_A\Pi \propto \mds{1}_B$, and $\Pi=\sum_i \Pi_i$ where $\Pi_i$ is a projector supported on $\mf{B}(AB_i)$ with $B=\bigoplus_i B_i$. If $\mcal{N}\otimes\id_B$ fixes $\Pi$, then it also fixes $\Pi_i$.
\end{prop}
\begin{proof}
    By applying $M=\sum_i \lambda_i \Pi_{\supp{B_i}}$ on $B$, with injective $i\mapsto \lambda_i$, we can transform $\Pi$ into $\sum_i \lambda_i \Pi_i$, so that they each $\Pi_i$ is a projector onto eigenspaces corresponding to a unique eigenvalue, and $\mcal{N}$ still preserves this operator. Thus the unital map $\mcal{N}\otimes\id_B$ should preserve it. Consequently, $\mcal{N}(\Tr_B\Pi_i)=\Tr_B\Pi_i.$
\end{proof}

\begin{prop} \label{prop:pcqkr}
    A quantum state $\rho_{AB}$ is PC-Q with respect to the essential decomposition $A=\bigoplus_i A_i$ (let $\Pi_i:=\mds{1}_{A_i}$) with corresponding type index sets $\mcal{I}_A$ and $\mcal{II}_A$ if and only if for any $M\in\mf{B}(A)$, $[M\otimes\mds{1}_B,\rho_{AB}]=0$ is equivalent to $M=\left(\bigoplus_{i\in \mcal{I}_A}\alpha_i\Pi_i\right)\oplus \left(\bigoplus_{i\in \mcal{II}_A}M_i\right)$ for some complex numbers $\alpha_i$ for all $i\in\mcal{I}_A$ and some $M_i\in\mf{B}(A_i)$ for all $i\in\mcal{II}_A$ (we will say that ``$M$ is in the standard form'').
\end{prop}
\begin{proof}
    Assume that $\rho_{AB}$ is a PC-Q state given as in the statement of Proposition. Let $M\in\mf{H}(A)$ commute with $\rho_{AB}$ but be not of the form $M=\left(\bigoplus_{i\in \mcal{I}_A}\Pi_i\right)\oplus \left(\bigoplus_{i\in \mcal{II}_A}M_i\right)$. It is impposible as it should be one of the following cases.
   
   
   $(i)$ $M$ is not block-diagonal with respect to the decomposition $A=\bigoplus_i A_i$: $M$ has the spectral decomposition $M=\sum_i m_i P_i$ ($m_i$ is unique for each $i$), but some $P_i$ does not commute with some of $\{\Pi_i\}$. Since $\rho_{AB}$ commutes with $M\otimes\mds{1}_B$, $\rho_{AB}$ is also generalized block-diagonal with respect to the eigenspaces of $M$, i.e., $[P_i\otimes\mds{1}_B,\rho_{AB}]=0$ for all $i$. It violates $(iii)$ of Definition \ref{def:CPQ}.
   
   $(ii)$ $M$ is block-diagonal with respect to the decomposition $A=\bigoplus_i A_i$, but for some $i\in \mcal{I}_A$, $\Pi_i M \Pi_i \not\propto \Pi_i$: We assume $[\Pi_i\otimes\mds{1}_B,\rho_{AB}]=0$ for all $i$ (See $(i)$ above). Let $M_i:=\Pi_i M \Pi_i$ and $\rho_i:=(\Pi_i\otimes \mds{1}_B)\rho_{AB}(\Pi_i\otimes\mds{1}_B)$. We have $[M_i\otimes\mds{1}_B,\rho_i]=0$, hence $\rho_i$ is PC-Q state with respect to the nontrivial eigenspaces of $M_i$. It violates $(ii)$ of Definition \ref{def:CPQ}.
   
   For general $M\in\mf{B}(A)$, one can consider its real and imaginary parts $M_R:=(M+M^\dag)/2$ and $M_I:=-i(M-M^\dag)/2$, which are Hermitian operators commuting with $\rho_{AB}$ themselves. The same argument applies to each of them, and the desired result follows for $M$: $A=\bigoplus_i A_i$ cannot be the essential decomposition of $A$ for $\rho_{AB}$.
   
    Likewise, let $M$ be given as $M=\left(\bigoplus_{i\in \mcal{I}_A}\Pi_i\right)\oplus \left(\bigoplus_{i\in \mcal{II}_A}M_i\right)$ but assume that $[M\otimes\mds{1}_B,\rho_{AB}]\neq0$. However, it is impossible if $A=\bigoplus A_i$ is the essential decomposition for $\rho_{AB}$ since type II components of $\rho_{AB}$ only can be in the form of $\pi_{A_i}\otimes \sigma_B$ for some $\sigma\in\mf{S}(B)$, so that $M\otimes\mds{1}_B$ of the standard form must commute with $\rho_{AB}$. It follows that $A=\bigoplus_i A_i$ cannot be the essential decomposition of $A$ for $\rho_{AB}$.

   Conversely, $A=\left(\bigoplus_{i\in\mcal{I}_A}A_i\right)\oplus\left(\bigoplus_{i\in\mcal{II}_A}A_i\right)$ be an arbitrary nontrivial decomposition of $A$ with $\Pi_i:=\mds{1}_{A_i}$. Assume that for any $M\in\mf{B}(A)$, $[M\otimes\mds{1}_B,\rho_{AB}]=0$ is equivalent to $M$ being in the standard form, i.e. $M=\left(\bigoplus_{i\in \mcal{I}_A}\alpha_i\Pi_i\right)\oplus \left(\bigoplus_{i\in \mcal{II}_A}M_i\right)$  for some complex numbers $\alpha_i$ for all $i\in\mcal{I}_A$ and some $M_i\in\mf{B}(A_i)$ for all $i\in\mcal{II}_A$. We check if the decomposition $A=\bigoplus_i A_i$ satisfies the three criteria of Definition \ref{def:CPQ}.
   
   $(i)$ Each $\Pi_i$ is obviously in the standard form, thus $[\Pi_i\otimes \mds{1}_B,\rho_{AB}]=0$. The decomposition is assumed to be nontrivial, thus no $\Pi_i$ is equal to $\mds{1}_A$.
   
   $(ii)$ For some $i\in\mcal{I}_A$, if $(\Pi_i\otimes \mds{1}_B)\rho_{AB}(\Pi_i\otimes \mds{1}_B)$ is PC-Q, then, it has its own essential decomposition of $A_i=\bigoplus_j A_i^j$, which can give a finer decomposition of $A$ and makes $\mds{1}_{A_i^j}\otimes \mds{1}_B$ commute with $\rho_{AB}$ even though $\mds{1}_{A_i^j}$ is not of the standard form. For some $i\in\mcal{II}_A$, if $(\Pi_i\otimes \mds{1}_B)\rho_{AB}(\Pi_i\otimes \mds{1}_B)\not\propto \mds{1}_{A_i}\otimes\sigma_B$ for some $\sigma\in\mf{S}(B)$, there exists some $W\in\mf{B}(A_i)$ such that $[W_A\otimes\mds{1}_B,\rho{AB}]\neq 0$ even though it is of the standard form.
   
   $(iii)$ Any projector $P$ on $A$ that does not commute with some $\Pi_i$ is not of the standard form, thus $[P\otimes\mds{1}_B,\rho_{AB}]\neq0$.

\end{proof}
Here, we provide a proof of Theorem \ref{thm:pcqunt}.\\
\textbf{Theorem \ref{thm:pcqunt}.} \;A unital channel $\mcal{N}\in\mf{UC}(A)$ fixes a quantum state $\rho_{AB}$ that is PC-Q with respect to the essential decomposition $A=\bigoplus_i A_i$ (let $\Pi_i:=\mds{1}_{A_i}$) with corresponding type index sets $\mcal{I}_A$ and $\mcal{II}_A$ if and only if $\mcal{N}$ preserves every subspace $A_i$ and acts trivially on $A_i$ when $i\in\mcal{I}_A$.

\begin{proof}
    If $\mcal{N}$ preserves every subspace $A_i$ and acts trivially on $\mf{B}(A_i)$ when $i\in\mcal{I}_A$, then $\mcal{N}\circ\Ad_{\Pi_i}=\Ad_{\Pi_i}$ for all $i$, so that $\mcal{N}\circ(\sum_i \Ad_{\Pi_i})=\sum_i \mcal{N}\circ\Ad_{\Pi_i}=\sum_i \Ad_{\Pi_i}$. Therefore, $\rho_{AB}$, which is fixed by $\sum_i \Ad_{\Pi_i}\otimes \id_B$, is also fixed by $\mcal{N}$.
    
    Conversely, assume that a unital channel $\mcal{N}$ fixes a PC-Q quantum state $\rho_{AB}$ with the structure given in the assumption. Hence, for every Kraus operator $K_j$ of $\mcal{N}$, i.e., $\mcal{C}=\sum_i \Ad_{K_j}$, we have $[K_j\otimes\mds{1}_B,\rho_{AB}]=0$ \cite{arias2002fixed,lie2020uniform,lie2021correlational}. By Proposition \ref{prop:pcqkr}, it follows that $K_j=\left(\bigoplus_{i\in \mcal{I}_A}\alpha_i^{(j)}\Pi_i\right)\oplus \left(\bigoplus_{i\in \mcal{II}_A}K_i^{(j)}\right)$ for some complex numbers $\alpha_i$ for all $i\in\mcal{I}_A$ and some $K_i^{(j)}\in\mf{B}(A_i)$ for all $i\in\mcal{II}_A$. By the trace preserving condition, $\sum_j K_j^\dag K_j =\mds{1}_A$, we have $\sum_j |\alpha_i^{(j)}|^2=1$. From the forms of $K_j$, we can see that $\mcal{C}=\left(\bigoplus_{i\in\mcal{I}_A} \id_{A_i}\right)\oplus\left(\bigoplus_{i\in\mcal{II}_A} \mcal{C}_i\right)$, with $\mcal{C}_i=\sum_j \Ad_{K_j^{(i)}}$ for every $i\in\mcal{II}_A$. It proves the desired result.
\end{proof}

\begin{lemma} \label{lem:fixing}
    For a unital channel $\mcal{N}\in\mf{UC}(A)$, if $\mcal{N}^\dag\circ\mcal{N}(\Pi_i)=\Pi_i$ for some partition of unity (i.e. $\{\Pi_i\}$ are projectors and $\sum_i\Pi_i=\mds{1}_A$), then there exists $U\in\mf{U}(A)$ and $\mcal{M}\in\mf{UC}(A)$ such that $\mcal{N}=\Ad_U\circ\mcal{M}$ and $\mcal{M}(\Pi_i)=\Pi_i$ for all $i$.
\end{lemma}
\begin{proof}
    As $\mcal{N}$ is unital, $\mcal{N}(\Pi_i)\prec\Pi_i$, but $\mcal{N}^\dag\circ\mcal{N}(\Pi_i)=\Pi_i$, i.e,  $S(\pi_i)=S(\mcal{N}(\pi_i))$ \cite{zhang2011neumann} where $\pi_i=|\Pi_i|^{-1}\Pi_i$. Because the von Neumann entropy is strictly Schur concave \cite{aniello2017quantum}, it means that there exists $U_i\in \mf{U}(A)$ such that $\mcal{N}(\Pi_i)=\Ad_{U_i}(\Pi_i)$. Hence $\mds{1}_A=\mcal{N}(\mds{1}_A)=\sum_i \mcal{N}(\Pi_i)=\sum_i \Ad_{U_i}(\Pi_i)$. From this we can deduce that $\sum_{i\neq j} \Pi_i = \sum_{i\neq j} \Ad_{U_j^\dag U_i}(\Pi_i)$. Thus, $\Tr[\Pi_j \Ad_{U_j^\dag U_i}(\Pi_i)]=\Tr[\Ad_{U_j}(\Pi_j)\Ad_{U_i}(\Pi_i)]=0$ for any $i$ and $j$. Hence $\{U_i \Pi_i U_i^\dag\}$ is another partition of unity with the same ranks, so there exists some unitary operator $V\in\mf{U}(A)$ such that $\Ad_{U_i}(\Pi_i)=\Ad_V(\Pi_i)$. It follows that $\mcal{M}:=\Ad_{V^\dag}\circ\mcal{N}$ is a unital channel on $A$ that preserves every $\Pi_i$ and $\mcal{N}=\Ad_V\circ\mcal{M}$.
\end{proof}

\begin{corollary} \label{coro:pcq}
     A unital channel $\mcal{N}\in\mf{UC}(A)$ does not increase the entropy of a quantum state $\rho_{AB}$ that is PC-Q with respect to the essential decomposition $A=\bigoplus_i A_i$ ($\Pi_i:=\mds{1}_{A_i}$) with corresponding type index sets $\mcal{I}_A$ and $\mcal{II}_A$ if and only if $\mcal{N}$ can be decomposed into $\mcal{N}=\Ad_V\circ\mcal{N}'$ with some unitary operator $V\in\mf{U}(A)$ and a unital channel $\mcal{N}'$ that preserves every subspace $A_i$ and acts trivially on $A_i$ when $i\in\mcal{I}_A$.
\end{corollary}
\begin{proof}
    By Theorem \ref{thm:pcqunt} and Lemma \ref{lem:fixing}, $\mcal{N}$ can be decomposed into $\mcal{N}=\Ad_{V'}\circ\mcal{M}$ with some $V'\in\mf{U}(A)$ and a unital channel $\mcal{M}\in\mf{UC}(A)$ that preserves every subspace $A_i$. Let $p_i:=\Tr[\Pi_i \rho_A]$ and $\rho_{AB}^i:=p_i^{-1}(\Ad_{\Pi_i}\otimes\id_B)(\rho_{AB})$. Then, there exists $\mcal{M}_i \in \mf{UC}(A_i)$ such that $(\mcal{M}\otimes\id_B)(\rho_{AB}^i)=(\mcal{M}_i\otimes\id_B)(\rho_{AB}^i)$ because each $\rho_{AB}^i$ is supported on $A_i\otimes B$ for every i. For $N$ to preserve the von Nuemann entropy of $\rho_{AB}$, it is required for every $\mcal{M}_i$ to preserve the von Neumann entropy of $\rho_{AB}^i$. For type I indices, i.e., when $i\in\mcal{I}_A$, it means that $\mcal{M}_i^\dag\circ\mcal{M}_i$ fixes $\rho_{AB}^i$ \cite{zhang2011neumann}. However, since $\rho_{AB}^i$ is not PC-Q as a state of $A_iB$, it follows that $\mcal{M}_i$ is a unitary operation on $A_i$ (See the proof of Theorem \ref{thm:delodep}), i.e., $\mcal{M}_i=\Ad_{W_i}$ for some $W_i\in\mf{U}(A_i)$. If we let $V:=V'R$ where $R:=\left(\bigoplus_{i\in\mcal{I}_A}W_i \oplus \bigoplus_{i\in\mcal{II}_A}\Pi_i\right)$ and $\mcal{N}':=\Ad_{R^\dag}\circ\mcal{M} $, then $\mcal{N}=\Ad_V \circ \mcal{N}'$ is the desired decomposition of $\mcal{N}$.
\end{proof}

\begin{corollary} \label{coro:pcpc}
     Let a delocalized catalysis unitary operator pair $(U_0,U_1)$ be compatible with a delocalized randomness source $\sigma_{B_0B_1}$ with the DCD
     \begin{equation}
         \sigma_{B_0B_1}=\bigoplus_{ij} (\Pi_i^{B_0}\otimes\Pi_j^{B_1})\sigma_{B_0B_1}(\Pi_i^{B_0}\otimes\Pi_j^{B_1}),
     \end{equation}
     and the essential decompositions $B_0=\bigoplus_i B_{0i}$ and $B_1=\bigoplus_i B_{1i}$. It follows that there exist $W_i\in\mf{U}(B_i)$ for $i=0,1$ ($\Pi_k^{B_i}:=\mds{1}_{B_{ik}}$) such that $U_i=(\mds{1}_{A_i}\otimes W_i)\left(\bigoplus_k U_{ik}\right)$ where $U_{ik}\in\mf{U}(A_iB_{ik})$ is a catalysis unitary operator compatible with $(\Pi_k^{B_0}\otimes\mds{1}_B)\sigma_{B_0B_1}(\Pi_k^{B_0}\otimes\mds{1}_B)$ for $i=0$ and $(\mds{1}_A\otimes\Pi_k^{B_1})\sigma_{B_0B_1}(\mds{1}_A\otimes\Pi_k^{B_1})$ for $i=1$.
\end{corollary}

\begin{proof}
    Consider the maximally mixed initial state $\pi_{A_0}\otimes\pi_{A_1}$ for the catalysis and let $\mcal{N}_i\in\mf{UC}(B_i)$ given as $\mcal{N}_i:=\mf{Tr}_{A_i}\circ\Ad_{U_i}$ be the induced catalytic map on $B_i$ acting on the delocalized catalyst $\sigma_{B_0B_1}$ for $i=0,1$. Since $(U_0,U_1)$ and $\sigma_{B_0B_1}$ are compatible with each other, we have $(\mcal{N}_0\otimes\mcal{N}_1)(\sigma_{B_0B_1})=\sigma_{B_0B_1}$. It follows that both of $\mcal{N}_i$ do not increase the entropy of $\sigma_{B_0B_1}$. By Corollary \ref{coro:pcq}, the Kraus operators $\{R_k^i\}$ of $\mcal{N}_i$ all have the form $R_k^i=W_i L_K^i$ where $W_i\in\mf{U}(B_i)$ is a unitary operator and $L_k^i$ is the Kraus operator of another unital map in the standard form. However, we recall that $K_{nm}^i:=|A_i|^{-1/2}(\bra{n}_{A_i}\otimes\mds{1}_{B_i})U_0(\ket{m}_{A_i}\otimes\mds{1}_{B_i})$ are the Kraus operators of $\mcal{N}_i$ for $i=0,1$. Let the standard form expression of $K_{nm}^i$ given as follows:
    \begin{equation}
        |A_i|^{1/2}K_{nm}^i=\left(\bigoplus_{k\in \mcal{I}_{B_i}}\alpha_k^{(i,n,m)}\Pi_k^{B_i}\right)\oplus \left(\bigoplus_{k\in \mcal{II}_{B_i}}K_k^{(i,n,m)}\right).
    \end{equation}
    Now, we let $U_{ik}:=\sum_{nm}\dyad{n}{m}_{A_i}\otimes \left(\alpha_k^{(i,n,m)}\Pi_k^{B_i}\right)$ for $k\in\mcal{I}_{B_i}$ and $U_{ik}:=\sum_{nm}\dyad{n}{m}_{A_i}\otimes K_k^{(i,n,m)}$ for $k\in\mcal{II}_{B_i}$ and the desired result follows.
\end{proof}

\subsection{Proof of Theorem \ref{thm:delext}}

\begin{proof}
For the given randomness source $\sigma_{B_0B_1}$, we again let $p_{ij}:=\Tr[(\Pi_i^{B_0}\otimes\Pi_j^{B_1})\sigma_{B_0B_1}]$ and $\sigma^{ij}:=p_{ij}^{-1}(\Ad_{\Pi_i^{B_0}}\otimes\Ad_{\Pi_j^{B_1}})(\sigma_{B_0B_1})$. We first show that it is achievable. To match the notations with (\ref{eqn:dlc1}) and (\ref{eqn:dlc2}), we set $\sigma_{B_0B_1}$ as our delocalized randomness source. (Consider that $A\to B_0$ and $B\to B_1$ in Definition \ref{def:CPQ} and \ref{def:DCD}.) We let $B_0=\bigoplus_i B_{0i}$ and $B_1=\bigoplus_i B_{1i}$ be their respective essential decompositions. For the sake of simplicity, assume that $\mcal{I}_{B_i}=\{1,2,\cdots,|\mcal{I}_{B_i}|\}$ and $\mcal{II}_{B_i}=\{|\mcal{I}_{B_i}|+1,\cdots,|\mcal{I}_{B_i}|+|\mcal{II}_{B_i}|\}$ for $i=0,1$. Also, let the dimension of $A_i$ be $|\mcal{I}_{B_i}|+\sum_{j\in\mcal{II}_{B_i}}|B_{ij}|^2$ for $i=0,1$. We use catalysis unitary operators
\begin{equation}
    U_0=\sum_{i\in\mcal{I}_{B_0}} Z_0^i\otimes \Pi_i^{B_0} +\!\! \sum_{i\in\mcal{II}_{B_0}}\!\!\sum_{j,k=0}^{|B_{0i}|-1}Z_0^{S_i^{0}+j|B_{0i}|+k}\otimes E_{jk}^{(0i)},
\end{equation}
and
\begin{equation}
    U_1=\sum_{i\in\mcal{I}_{B_1}} Z_1^i\otimes \Pi_i^{B_1} +\!\! \sum_{i\in\mcal{II}_{B_1}}\!\!\sum_{j,k=0}^{|B_{1i}|-1}Z_1^{S_i^{1}+j|B_{1i}|+k}\otimes E_{jk}^{(1i)}.
\end{equation}
Here, $E_{jk}^{(im)}:=\omega_{im}^{jk} \dyad{m_j^i}{m_k^i}$, where $\omega_{im}$ is the $|B_{im}|$-th root of the unity, and $\{\ket{m_j^i}\}$ is an orthonormal basis of $B_{im}$.  are an arbitrary orthonormal unitary operator on $B_{0i}$ and $B_{1i}$, respectively. Also, $Z_i=\sum_k \dyad{k\oplus1 (\text{mod }|A_i|)}{k}$ is the generalized Pauli-Z operator on $A_i$, and $S_k^i:=|\mcal{I}_{B_i}|+\sum_{l=0}^{k-1}|B_{il}|^2$ assuming $|B_{i0}|=0$ for $i=0,1$ . Now, we suppose that system $A_0A_1$ is prepared in a maximally entangled state $\ket{\phi^+}_{A_0A_0'}\ket{\phi^+}_{A_1A_1'}$ with auxiliary systems $A_0'$ and $A_1'$. After the catalysis, the final state of $A_0A_1A'_0A'_1$ is
\begin{equation} \label{eqn:catout}
\begin{split}
    &\sum_{\substack{i\in\mcal{I}_{B_0},\\j\in\mcal{I}_{B_1}}} p_{ij} \phi_{A_0A'_0}^{i}\otimes\phi_{A_1A'_1}^{j} + \sum_{\substack{i\in\mcal{I}_{B_0},\\j\in\mcal{II}_{B_1}}} p_{ij} \phi_{A_0A'_0}^{i}\otimes\Xi_j^1 \\
    +&\sum_{\substack{i\in\mcal{II}_{B_0},\\j\in\mcal{I}_{B_1}}} p_{ij} \Xi_i^0\otimes\phi_{A_1A'_1}^{j}
   +\sum_{\substack{i\in\mcal{II}_{B_0},\\j\in\mcal{II}_{B_1}}} p_{ij} \Xi_i^0\otimes\Xi_j^1.
\end{split}
\end{equation}
Here, $\phi_{A_iA'_i}^m:=\Ad_{Z_i^{m}}\otimes\id_{A'_i}(\phi_{A_iA'_i}^+)$ are mutually orthogonal Bell states for $i=0,1$. Also, $\Xi_i^0$ and $\Xi_j^1$ are given as
\begin{equation}
    \Xi_i^0:=\frac{1}{|B_{0i}|^2}\sum_{k=0}^{|B_{0i}|^2-1}{\phi_{A_0A'_0}^{S_i^0+k}},
\end{equation}
and
\begin{equation}
    \Xi_j^1:=\frac{1}{|B_{1j}|^2}\sum_{l=0}^{|B_{1j}|^2-1}{\phi_{A_1A'_1}^{S_j^1+l}},
\end{equation}
for all $i\in\mcal{II}_{B_0}$ and $j\in\mcal{II}_{B_1}$. Note that $\Xi_i^0$ and $\Xi_j^1$ are unitarily similar with $\pi_{\mds{C}^{|B_{0i}|^2}}$ and $\pi_{\mds{C}^{|B_{1j}|^2}}$, respectively. Since every term in (\ref{eqn:catout}) is mutually orthogonal to each other, it is unitarily similar to $\bigoplus_{i,j} p_{ij} \tau_i\otimes\kappa_j$ in (\ref{eqn:DCE}), after the changes of labels.

Conversely, by Corollary \ref{coro:pcpc}, every component in the DCD of a delocalized catalyst is compatible up to local unitary with the given pair of catalysis unitary operators by itself. Let $\mcal{C}$ be the catalytic map implemented by the catalysis unitary operators $U_0$ and $U_1$ by using $\sigma_{B_0B_1}$ as the catalyst. In other words, $\mcal{C}(\rho):=\Tr_{B_0B_1}[(\Ad_{U_0}\otimes\Ad_{U_1})(\rho_{A_0A_1}\otimes\sigma_{B_0B_1})]$. Now we let $\mcal{C}_{ij}$ be given as $\mcal{C}_{ij}(\rho):=\Tr_{B_0B_1}[(\Ad_{U_0}\otimes\Ad_{U_1})(\rho_{A_0A_1}\otimes\sigma_{B_0B_1}^{ij})]$, which is a catalytic map by itself, then we have $\mcal{C}=\sum_{ij}p_{ij}\mcal{C}_{ij}$. For arbitrary pure initial state $\rho_{A_0A_1}$ (recall that the maximum entropy production is made with a pure state input), we have the following.
\begin{equation}
\begin{split}
    \mcal{C}(\rho)&=\sum_{ij}p_{ij}\mcal{C}_{ij}(\rho)\succ\bigoplus_{ij}p_{ij}\mcal{C}_{ij}(\rho)\\
    &\succ\bigoplus_{ij} p_{ij} \tau_i\otimes\kappa_j.
\end{split}
\end{equation}
The first majorization relation follows from the fact that a convex sum of quantum states always majorizes the direct sum of the same summands \cite{nielsen2000probability}. The last majorization relation follows because whenever $k\in\mcal{I}_{B_i}$, $U_i$ can only act unitarily on $B_{ik}$, hence no randomness can be extracted on that side, and when $l\in\mcal{II}_{B_i}$, then the catalyst is in a product state in that component, thus it simply functions as a single party randomness source. It means that $\tau_l$ (or $\kappa_l$) functions as the REO. We also used the fact that a direct sum of quantum states majorizes another when its individual summand majorizes that of the other. Since every \renyi entropy of order $\alpha\geq0$ is Schur-concave, the desired result follows.
\end{proof}
\subsection{Proof of Theorem \ref{thm:semthm}}
Let us first show that utilization of semantic information is a special case of randomness utilization.
\begin{lemma} \label{lem:sdync1}
    If $U\in\mf{U}(AB)$ and $\sigma_{AB}\in\mf{S}(AB)$ are given as in Definition \ref{def:seminf}, then $U$ is a catalysis unitary operator compatible with $\sigma_B$ as a catalyst up to local unitary.
\end{lemma}
\begin{proof}
    As any superchannel can be decomposed into pre- and post-processing channels, (\ref{eqn:semcat}) is equivalent to 
    \begin{equation} \label{eqn:semcat2}
    \Tr_A[\,\mcal{U}\circ(\mcal{N}_A\otimes\id_B)(\sigma_{AB})]=\eta_B,
    \end{equation}
    for any channel $\mcal{N}\in\mf{C}(A)$. Here, $\mcal{N}$ is the partial trace of the arbitrarily chosen pre-processing channel of $\Theta_{A\to B}$ in (\ref{eqn:semcat}). By letting $\mcal{N}$ be a state preparation channel, i.e. $\mcal{N}(\rho)=\tau_A\Tr\rho$ for every $\tau\in\mf{S}(A)$, we get that
    \begin{equation} \label{eqn:altdef}
    \Tr_A[\,\mcal{U}(\tau_A\otimes\sigma_B)]=\eta_B,
    \end{equation}
    for any $\tau\in\mf{S}(A)$. By the result of Ref. \cite{lie2020uniform}, there exists a unitary operator $V$ such that $\eta_B = \Ad_V(\sigma_B)$, thus by the definition given in (\ref{eqn:catal2}), $U$ is a catalysis unitary operator and it is compatible with $\sigma_B$ up to local unitary.
\end{proof} \label{prop:supcha}
As a side note, this Lemma provides a proof of the first part of Theorem \ref{thm:picinv}. That is, if $\sigma_{AB}$ is uncorrelated, i.e., $\sigma_{AB}=\sigma_A\otimes\sigma_B$, then every catalytic unitary operation compatible with $\sigma_B$ as a catalyst utilizes only information of $B$ in $\sigma_{AB}$. It is because if $\sigma_{AB}=\sigma_A\otimes\sigma_B$, then  (\ref{eqn:semcat2}) becomes equivalent to
    \begin{equation}
        \Tr_A[\,\mcal{U}(\rho_A\otimes\sigma_B)]=\sigma_B,
    \end{equation}
    for every $\rho\in\mf{S}(A)$ as the set $\{\mcal{N}(\sigma_A)\,|\,\mcal{N}\in\mf{C}(A)\}$ is same with $\mf{S}(A)$. Since it is equivalent to (\ref{eqn:catal2}), we get the desired result.

((S:B) $\Rightarrow$ (S:A)) It immediately follows from the fact that any superchannel can be decomposed into pre- and post- processes. Note that the output of the transformed channel on $A$ is immediately discarded, the post-process is irrelevant. The process $\mcal{N}_{A\to RA}$ can be considered the pre-process of the superchannel $\Theta$ in (S:A).

((S:C) $\Leftrightarrow$ (S:B)) Without loss of generality, we consider the canonical case (without local unitary transformation on catalysts), if $U\in\mf{U}(AB)$ is compatible with $\sigma_{AB}$ on $B$, we have
\begin{equation}
    \Tr_{A'}\circ \Ad_{U_{A'B}} (\sigma_{AB}) = \Tr_{A'} \otimes \sigma_{AB}.
\end{equation}
A simple change of system labels yields that for every $\mcal{L}\in\mf{L}(A)$ (by considering it as linear map that maps from $A$ to $A'$), we have
\begin{equation}
    \Tr_{A}\circ \Ad_{U_{AB}} \circ \mcal{L}_A (\sigma_{AB}) = \Tr_{A} \circ \mcal{L}_A( \sigma_{AB}).
\end{equation}
By inserting arbitrary quantum map $\mcal{N}\in\mf{C}(A,RA)$ into the position of $\mcal{L}_A$, we have the desired result
\begin{equation}
    \Tr_{A}\circ \Ad_{U_{AB}} \circ \mcal{N}_{A\to RA} (\sigma_{AB}) = \Tr_{A} \circ \mcal{N}_{A\to RA}( \sigma_{AB}).
\end{equation}

By choosing $\mcal{N}_{A\to RA}= \dyad{\psi}_A \otimes \id_{A\to R}$ for each state $\ket{\psi}$ on $A$, one can also show the converse.

((S:A) $\Rightarrow$ (S:C)) We will use the following Lemma.
\begin{lemma} \label{lem:conssup}
    For any constant superchannel $\Theta$ that maps channels in $\mf{C}(A,B)$ to channels in $\mf{C}(C,D)$, meaning that $\Theta(\mcal{N})$ is same for every $\mcal{N}\in\mf{C}(A,B)$, there exists a quantum channel $\mcal{P}\in\mf{C}(C,AD)$ such that
    \begin{equation}
        \Theta(\mcal{L})= (\Tr_B\circ\mcal{L}_{A\to B}\otimes\id_D)\circ \mcal{P}_{C\to AD},
    \end{equation}
    for any $\mcal{L}\in\mf{L}(A,B)$.
\end{lemma}
\begin{proof}
A basis of $\mf{L}(A,B)$ is $\{\mcal{E}_{ij}:=Y_j\Tr[X_i^\dag\;\cdot\;]\},$ where $\{X_i\}$ and $\{Y_j\}$ are orthonormal basis of $\mf{B}(A)$ and $\mf{B}(B)$ respectively that consist of traceless Hermitian operators except for $X_0=|A|^{-1/2}\mds{1}_A$ and $Y_0=|B|^{-1/2}\mds{1}_B$. Hence, every $\mcal{L}\in\mf{L}(A,B)$ has the expression of the following form,
\begin{equation}
    \mcal{L}=\sum_{ij} \mcal{E}_{ij} \Tr[Y_j^\dag \mcal{L}(X_i)].
\end{equation}
Note that the span of $\mf{C}(A,B)$ coincides with the span of $\{\mcal{E}_{ij}\}$ excluding $\mcal{E}_{i0}$ with $i>0$. If we let $\mcal{F}_{ij}:=\Theta(\mcal{E}_{ij})\in\mf{L}(C,D)$, we get the expression
\begin{equation}
    \Theta(\mcal{L})=\sum_{ij} \mcal{F}_{ij} \Tr[Y_j^\dag\mcal{L}(X_i)].
\end{equation}
By the condition that $\Theta$ is constant for quantum channels in $\mf{C}(A,B)$, there exists some channel $\mcal{C}\in\mf{C}(C,D)$ such that $\Theta(\mcal{N})=\mcal{C}$ for all $\mcal{N}\in\mf{C}(A,B)$ and
\begin{equation} \label{eqn:compa1}
    \Theta(\mcal{L})=\mcal{C}\Tr[\mcal{L}(\pi_A)]+\sum_{i>0} \mcal{F}_{i0} \Tr[\mcal{L}(X_i)].
\end{equation}
Now, we let $\mcal{P}\in\mf{L}(C,AD)$ defined as
\begin{equation} \label{eqn:compa2}
    \mcal{P}:=\pi_A\otimes \mcal{C} + \sum_{i>0}X_i\otimes \mcal{F}_{i0}.
\end{equation}
From (\ref{eqn:compa1}), we can see that if $\mcal{Q}\in\mf{C}(C,AE)$ and $\mcal{R}\in\mf{C}(BE,D)$ are pre- and post-processing channels of $\Theta$ so that $\Theta(\mcal{L})=\mcal{R}\circ(\mcal{L}\otimes\id_E)\circ\mcal{Q}$ for every $\mcal{L}\in\mf{L}(A)$, then $\mcal{P}_{C\to AD}=(\mcal{R}_{A'E\to D}\otimes\id_A)(\tau_{A'}\otimes \mcal{Q}_{C\to AE})$ for some $\tau\in\mf{S}(A')$. Therefore, as a composition of quantum channels, $\mcal{P}$ is obviously a quantum channel. Moreover, by comparing (\ref{eqn:compa1}) and (\ref{eqn:compa2}) , we get the desired result
\begin{equation}
        \Theta(\mcal{L})= (\Tr_B\circ\mcal{L}_{A\to B}\otimes\id_D)\circ \mcal{P}_{C\to AD}.
    \end{equation}
\end{proof}

Indeed, as we can observe that the left hand side of (\ref{eqn:semcat2}) is a constant superchannel when $\mcal{N}$ is considered an input, we can apply Lemma \ref{lem:conssup}. Therefore, there exists a quantum state (which is a special type of quantum channel) $\tau_{AB}$ such that
\begin{equation}
     \Tr_A[ \,\mcal{U}\circ(\mcal{L}_A\otimes\id_B)(\sigma_{AB})]=\Tr_A[(\mcal{L}_A\otimes\id_B)(\tau_{AB})],
\end{equation}
for every $\mcal{L}\in\mf{L}(A)$. Equivalently, inputting a part of the swapping gate on $AA'$, we get
\begin{equation}\label{eqn:ranut}
     \Tr_{A'}[ \,(\mcal{U}_{A'B}\otimes \id_A)(\rho_{A'}\otimes\sigma_{AB})]=\tau_{AB},
\end{equation}
for all $\rho_{A'}\in\mf{S}(A')$. In other words, the mapping $\rho_{A'}\mapsto \tau_{AB}$ is constant. If one interpret (\ref{eqn:ranut}) as that $U_{A'B}\otimes\mds{1}_A$ utilizes $\sigma_{AB}$ as a randomness source, by the result of Ref. \cite{lie2020uniform}, $\tau_{AB}$ must have the same spectrum, thus also the same entropy, with $\sigma_{AB}$. Then, by Corollary \ref{coro:pcq}, there exists a unitary operator $V\in\mf{U}(B)$ such that $\tau_{AB}=\id_A\otimes\Ad_V(\sigma_{AB})$. This proves the desired result.

\subsection{Proof of Corollary \ref{coro:seminf}}
Let $\mcal{U}:=\Ad_U$. We will use the following Lemma.
\begin{lemma}[\cite{hayden2004structure}] \label{lem:vandi}
    If a tripartite state $\rho_{RAB}$ satisfies $I(R:A)=I(R:AB)$, then, the Hilbert space of $A$ has a direct sum structure of the form of $A=\bigoplus_i A_{i,K}\otimes A_{i,L}$ and $\rho_{RAB}$ can be decomposed into
    \begin{equation} \label{eqn:vandi}
        \rho_{RAB}=\bigoplus_i p_i \rho_{RA_{i,K}}\otimes\rho_{A_{i,L}B},
    \end{equation}
    where for each $i$, $\rho_{RA_{i,K}} \in R\otimes A_{i,K}$ and $\rho_{RA_{i,L}B}\in A_{i,L}\otimes B$. Additionally, it is equivalent to that $I(A:B)=I(RA:B)$.
\end{lemma}
By Lemma \ref{lem:vandi}, $\rho_{RAB}$ has the form of (\ref{eqn:vandi}). Therefore, its marginal state on $AB$ must have a form of
 \begin{equation} \label{eqn:vanma}
        \rho_{AB}=\bigoplus_i p_i \rho_{A_{i,K}}\otimes\rho_{A_{i,L}B}.
    \end{equation}
Since each subspace $A_{i,K}\otimes A_{i,L}$ is orthogonal to each other, we can construct quantum channels $\mcal{N}_i \in \mf{C}(A_{i,K},RA_{i,K}) $ such that $\mcal{N}_i(\rho_{A_{i,K}})=\rho_{RA_{i,K}}$. Therefore there exists a quantum map $\mcal{N}:=\bigoplus_i \mcal{N}_i\otimes \id_{A_{i,L}}\in \mf{C}(A,RA)$ that maps $\rho_{AB}$ into $\rho_{RAB}$.

\subsection{Proof of Theorem \ref{thm:leasod}}

\begin{proof}
     The essential decomposition of $\sigma_{AB}$ on $B$ has the following form.
    \begin{equation}
        \sigma_{AB} =\sum_{i\in \mcal{I}_B} p_i \sigma_{AB}^i + \sum_{i\in \mcal{II}_B} \sigma_A^i\otimes \sigma_B^i.
    \end{equation}
    The marginal state of $A$ after a general information utilization of $B$ has the following form.
    \begin{equation}
        \sum_{i\in \mcal{I}_B} p_i\Ad_{V_i}(\sigma_A^i)  +\sum_{i\in \mcal{II}_B} p_i \Phi_i(\sigma_A^i),
    \end{equation}
    where $\Phi_i$ are some catalytic maps on $A$ and $V_i \in \mf{U}(A)$. Since unitary operations are a special case of catalytic maps, one can simplify the expression and get
    \begin{equation}
        \sum_i p_i \Phi_i(\sigma_A^i).
    \end{equation}
    We claim that the probability distribution $\left(\sum_i p_i \lambda_j (\sigma_A^i)\right)$ majorizes $\left(\lambda_j(\sum_i p_i  (\sigma_A^i))\right)$. This is because of Fan's Lemma \cite{zhan2013matrix}.
    \begin{equation}
        \begin{split}
            \sum_{1\leq j \leq k} \lambda_j\left(\sum_i p_i \Phi_i(\sigma_A^i)\right) &= \max_{P}\Tr[P\sum_i p_i \Phi_i(\sigma_A^i)]\\
            = \max_{P}\Tr[\sum_i p_i P \Phi_i(\sigma_A^i)] &\leq  \max_{P}\sum_i p_i \Tr[P \Phi_i(\sigma_A^i)],\\
        \end{split}
    \end{equation}
    where the maximization is over rank-$k$ projectors $P$. Again by using Fan's Lemma \cite{zhan2013matrix}, we get
    \begin{equation}
        \sum_{1\leq j \leq k} \lambda_j\left(\sum_i p_i \Phi_i(\sigma_A^i)\right) \leq \sum_i p_i \sum_{1\leq j \leq k} \lambda_j(\Phi_i(\sigma_A^i)).
    \end{equation}
    From the relation between unital maps and majorization, we have $\Phi_i(\sigma_A^i)\succ \sigma_A^i$ for all $i$, hence $\sum_{1\leq j \leq k} \lambda_j(\Phi_i(\sigma_A^i) \leq \sum_{1\leq j \leq k} \lambda_j(\sigma_A^i)$ for all $i$ and $k$. Therefore, it follows that
    \begin{equation}
        \sum_{1\leq j \leq k} \lambda_j\left(\sum_i p_i \Phi_i(\sigma_A^i)\right) \leq \sum_i p_i \sum_{1\leq j \leq k} \lambda_j(\sigma_A^i),
    \end{equation}
    for all $k$. By choosing each $\Phi_i$ as a unitary operation that transforms $\sigma_A^i$ into $\sum_j\lambda_j(\sigma_A^i) \dyad{j}$ for some common basis $\{\ket{i}\}$, the catalytic transformation of $\sigma_A$ into $\sum_j (\sum_ip_i \lambda_j(\sigma_A^i)) \dyad{j}$ is achievable.
    
\end{proof}

\bibliography{main}
\end{document}